\documentclass[12pt]{article}
\pdfoutput=1
\usepackage{jheppub}
\usepackage{amsmath}
\usepackage{amsfonts}
\usepackage{amssymb}
\usepackage{graphicx}
\usepackage{mathrsfs}

\usepackage[export]{adjustbox}
\newcommand{\bmat}{\left(\begin{array}}
\newcommand{\emat}{\end{array}\right)}

\def\yzero{\smash{\hbox{$y\kern-4pt\raise1pt\hbox{${}^\circ$}$}}}

\def\beq{\begin{equation}}
\def\eeq{\end{equation}}
\def\beqa{\begin{eqnarray}}
\def\eeqa{\end{eqnarray}}

\def\-{\hphantom{-}}
\def\ov{\overline}
\def\s2{\frac{1}{\sqrt2}}

\def\beq{\begin{equation}}
\def\eeq{\end{equation}}
\def\beqa{\begin{eqnarray}}
\def\eeqa{\end{eqnarray}}

\def\IF{\relax{\rm I\kern-.18em F}}
\def\II{\relax{\rm I\kern-.18em I}}

\def\Dsl{\,\raise.15ex\hbox{/}\mkern-13.5mu D} %this one can be subscripted
\def\IS{{\bf {S}}}

\def\IZ{{\bf {Z}}}
\def\IX{{\bf {X}}}

\def\NN{{\cal {N}}}

\catcode`\@=11   
\newdimen\@rotdimen
\newbox\@rotbox  

\def\@vspec#1{\special{ps:#1}}%  passes #1 verbatim to the output
\def\@rotstart#1{\@vspec{gsave currentpoint currentpoint translate
   #1 neg exch neg exch translate}}% #1 can be any origin-fixing transformation
\def\@rotfinish{\@vspec{currentpoint grestore moveto}}% gets back in synch 
\def\@rotr#1{\@rotdimen=\ht#1\advance\@rotdimen by\dp#1%
   \hbox to\@rotdimen{\hskip\ht#1\vbox to\wd#1{\@rotstart{90 rotate}%
   \box#1\vss}\hss}\@rotfinish}
\def\@rotl#1{\@rotdimen=\ht#1\advance\@rotdimen by\dp#1%
   \hbox to\@rotdimen{\vbox to\wd#1{\vskip\wd#1\@rotstart{270 rotate}%
   \box#1\vss}\hss}\@rotfinish}%
\def\@rotu#1{\@rotdimen=\ht#1\advance\@rotdimen by\dp#1%
   \hbox to\wd#1{\hskip\wd#1\vbox to\@rotdimen{\vskip\@rotdimen
   \@rotstart{-1 dup scale}\box#1\vss}\hss}\@rotfinish}%
\def\@rotf#1{\hbox to\wd#1{\hskip\wd#1\@rotstart{-1 1 scale}%
   \box#1\hss}\@rotfinish}%
\def\rotate{\@ifnextchar[{\@rotate}{\@rotate[l]}}
\def\@rotate[#1]#2{\setbox\@rotbox=\hbox{#2}\@nameuse{@rot#1}\@rotbox}

\catcode`\@=12
\begin{document}

%----------------------------------------------------------------------%
%  numbering equations with section number
%----------------------------------------------------------------------%
\makeatletter
\@addtoreset{equation}{section}
\makeatother
\renewcommand{\theequation}{\thesection.\arabic{equation}}
%----------------------------------------------------------------------%
%  title page
%----------------------------------------------------------------------%
\pagestyle{empty}
%\vspace*{0.5in}
\rightline{IFT-UAM/CSIC-23-31}
\vspace{0.5cm}
\begin{center}
\Large{\bf Small Black Hole Explosions}
\\%[8mm] 
%}\\

\large{Roberta Angius, Jes\'us Huertas,  Angel M. Uranga\\[4mm]}
\footnotesize{Instituto de F\'{\i}sica Te\'orica IFT-UAM/CSIC,\\[-0.3em] 
C/ Nicol\'as Cabrera 13-15, 
Campus de Cantoblanco, 28049 Madrid, Spain}\\ 
\footnotesize{\href{roberta.angius@csic.es}{roberta.angius@csic.es},  \href{mailto: j.huertas@csic.es}{ j.huertas@csic.es},  \href{mailto:angel.uranga@csic.es}{angel.uranga@csic.es}}

\vspace*{10mm}

\small{\bf Abstract} \\%[5mm]
\end{center}
\begin{center}
\begin{minipage}[h]{\textwidth}
Small black holes are a powerful tool to explore infinite distances in moduli spaces.
However, we show that in 4d theories with a scalar potential growing fast enough at infinity, it is energetically too costly for scalars to diverge at the core, and the small black hole puffs up into a regular black hole, or follows a runaway behaviour.

We derive a critical exponent characterizing the occurrence or not of such small black hole explosions, both from a 4d perspective, and in the 2d theory after an $\IS^2$ truncation. The latter setup allows a unified discussion of fluxes, domain walls and black holes, solving an apparent puzzle in the expression of their potentials in the 4d $\NN=2$ gauged supergravity context.

We discuss the realization of these ideas in 4d $\NN=2$ gauged supergravities. Along the way we show that many regular black hole supergravity solutions in the literature in the latter context are incomplete, due to Freed-Witten anomalies (or duals thereof), and require the emission of strings by the black hole. 

From the 2d perspective, small black hole solutions correspond to dynamical cobordisms, with the core describing an end of the world brane. Small black hole explosions represent obstructions to completing the dynamical cobordism. We study the implications for the Cobordism Distance Conjecture, which states that in any theory there should exist dynamical cobordisms accessing all possible infinite distance limits in scalar field space. The realization of this principle using small black holes leads to non-trivial constraints on the 4d scalar potential of any consistent theory; in the 4d $\NN=2$ context, they allow to recover from a purely bottom-up perspective, several non-trivial properties of vector moduli spaces near infinity familiar from CY$_3$ compactifications.
\end{minipage}
\end{center}
\newpage
%----------------------------------------------------------------------%
%  Resetting of counters
%----------------------------------------------------------------------%
\setcounter{page}{1}
\pagestyle{plain}
\renewcommand{\thefootnote}{\arabic{footnote}}
\setcounter{footnote}{0}
%----------------------------------------------------------------------%
%  Paper begins
%----------------------------------------------------------------------%

\tableofcontents

\vspace*{1cm}

\newpage

\section{Introduction}

The exploration of regions at infinite distance in moduli spaces (or in general, scalar field spaces) is a powerful source of new physics results. One instance is the celebrated swampland distance conjecture \cite{Ooguri:2006in}\footnote{See e.g. \cite{Grimm:2018ohb,Corvilain:2018lgw,Grimm:2019ixq} for recent developments in CY moduli spaces.}, derived  from adiabatic motion in exact moduli spaces. However, a physically more realistic avenue for the exploration of infinite distances in field space, pioneered in \cite{Buratti:2018xjt} (see also \cite{Lust:2019zwm}), is the use of spacetime dependent solutions in effective field theories, in which certain scalars run along some spacetime coordinate.  

In several setups, such solutions involve scalars which run off to infinite distance at a finite distance in spacetime.  The resulting singularities describe physical objects whose properties relate to the infinite distance in interesting ways. In particular, the end of the world branes in dynamical cobordisms \cite{Buratti:2021yia,Buratti:2021fiv,Angius:2022aeq,Blumenhagen:2022mqw,Blumenhagen:2023abk} (see also \cite{McNamara:2019rup} for the swampland cobordism conjecture, \cite{Dudas:2000ff,Blumenhagen:2000dc,Dudas:2002dg,Dudas:2004nd} for early work and \cite{Basile:2018irz,Antonelli:2019nar,Mininno:2020sdb,Basile:2020xwi,Basile:2021mkd,Mourad:2022loy} for other related recent developments), display interesting scaling relations among the spacetime curvature and the spacetime and field theory distances. Other setups are 4d EFT strings \cite{Lanza:2020qmt,Lanza:2021qsu,Marchesano:2022avb}, and small black holes  (see e.g. \cite{Dabholkar:2012zz} for a review, and \cite{Hamada:2021yxy} for discussion in the swampland context). In fact, they can also be regarded as dynamical cobordisms of the theories upon compactification on $\IS^1$ for EFT strings and $\IS^2$ for 4d small black holes.

An important advantage of dynamical explorations of infinity in field space, as compared with the adiabatic one, is that the latter may be obstructed in the presence of non-trivial scalar potentials growing at infinity\footnote{In fact, taking fixed vevs in the slope of a non-trivial potential leads to dynamical tadpoles, which besides being unaesthetical, can also lead to specific incompatibilities with swampland constraints \cite{Mininno:2020sdb}.}, as emphasized in \cite{Gonzalo:2018guu}. In the context of ETW branes, the introduction of non-trivial potentials growing at infinity was included in the analysis in \cite{Angius:2022aeq}. In the present work, we deal with this question for small black holes. For concreteness, we focus on 4d small black holes, but we expect similar conclusions to apply in other dimensions.

Small black holes are very robust solutions in theories with $U(1)$'s and scalar-dependent gauge kinetic functions. Extremal small black holes may be regarded as the endpoint of the evaporation of large black holes with some charge. In addition, as emphasized in \cite{Hamada:2021yxy}, they are interesting probes of infinity in scalar moduli space\footnote{See \cite{Delgado:2022dkz} for a different exploration of infinity, using {\em large} black holes.}. As you approach the black hole core, a black hole potential drives the scalars towards infinite distance in field space, in a most radical realization of the attractor mechanism. Indeed, the near-core solution is independent of the value of moduli in the asymptotic vacuum infinitely away from the black hole. 

One could even expect a similar phenomenon, even in the presence of a 4d potential fixing the asymptotic moduli; the 4d scalar potential would be irrelevant near the black hole core, still allowing scalars to attain infinity in field space. 
However, this may not be so if the potential grows fast enough near infinity in field space, so that it is able to drag the scalars back from infinity. In such case, the small black hole core explodes\footnote{The name evokes the similar (but converse) effect in \cite{Green:2006nv}, where a black hole attractor drags the 4d scalar vevs from an asymptotic vacuum to another. Here we have a 4d potential dragging the scalar from the infinite values at the small black hole core to some finite values at a regular horizon. Both setups deal with the competition of black hole attractor potentials and 4d potentials.}, possibly puffing up into a black hole solution of finite size horizon, or perhaps following a runaway with no static endpoint configuration. 

In this paper we provide explicit examples of the different possibilities, and give precise criteria on the 4d potential for each of them. In particular, we find that for small black holes exploring an infinite distance limit (with gauge coupling scaling as a decreasing exponential along it), the small black hole explosion occurs when the 4d potential grows faster (or equal) than a critical exponent, related to that of the gauge coupling, but independent of the black hole charge. 

We study the implications of small black hole explosions for the capability of small black holes to explore infinite distance limits in scalar moduli spaces. Our findings motivate interesting non-trivial constraints on the growth near infinity of 4d scalar potentials in any consistent theory of quantum gravity. In fact, in 4d $\NN=2$ theories we obtain constraints on the behaviour of the K\"ahler potential near infinity in moduli space, matching, from a fully bottom-up approach, non-trivial properties of CY$_3$ compactifications.

We carry out the analysis both in the 4d picture, as well as in terms of the 2d truncation on the angular $\IS^2$. The latter viewpoint is interesting in itself \cite{Angius:2022aeq}, since the spacetime dependent solutions correspond to dynamical cobordisms, i.e. running configurations (for the scalars and the size of $\IS^2$), with the small black hole cores corresponding to the ETW branes, i.e. the cobordism defect required to absorb the charges. From this perspective, small black holes puffed up into regular black holes correspond to running solutions, which, instead of hitting an ETW brane which ends spacetime, relax to a minimum of the effective potential describing the near horizon AdS$_2$ vacuum. They represent solutions which fail to realize dynamical cobordisms ending spacetime. In this respect, they provide a large class of realizations of the idea, introduced in \cite{Angius:2022aeq} (in the context of  the D2/M2-brane solution \cite{Itzhaki:1998dd}), that crouching corrections in the near core region of an ETW brane can reveal hidden vacua coming down from infinity in moduli space.

Along the way in this work, we find several other interesting results. We show that this 2d perspective is optimal to discuss the interplay of domain walls, black holes and fluxes recently reviewed in \cite{Cribiori:2023swd} in the context of in 4d $\NN=2$ (gauged) supergravities. Moreover, our 2d perspective explains that the seemingly different expression for the black hole and the 4d potentials in terms of the covariantly holomorphic quantities, actually agree upon the inclusion of an extra field associated to the $\IS^2$ size (see \cite{Cribiori:2023swd} for an alternative proposed explanation). We also use a 2d version the entropy function formalism \cite{Sen:2005wa} (see \cite{Sen:2007qy} for a review) to provide a novel derivation of the effective potential in \cite{Bellucci:2008cb}, controlling the attractor mechanism in the presence of both black hole and 4d scalar potentials.

Although our discussion is general, a particular setup in which to study the competition of black hole attractor potentials and 4d potentials is 4d $\NN=2$ gauged supergravity. 
Indeed, regular black holes which feature a balance of both kinds of potentials have appeared in the literature, including BPS solutions \cite{Cacciatori:2009iz,DallAgata:2010ejj,Hristov:2010ri}, and non-supersymmetric solutions \cite{Klemm:2012yg,Klemm:2012vm,Gnecchi:2012kb} (see \cite{Astesiano:2021hro} and references therein for a recent discussion). However,  we argue that,  upon a closer look, there is a seemingly unnoticed consistency condition in general not satisfied by these solutions. The problem is related to the non-invariance of the configurations under some symmetries underlying Abelian gaugings, and in certain microscopic setups shows up as a Freed-Witten anomaly. The configurations can thus be amended by having the black hole emit a number of 4d strings (or equivalently having those strings fill up the near horizon AdS$_2$). For regular black holes with large charges $Q$, the anomaly is a $1/Q$ effect, meaning that these strings can be described as probes in the above classical geometries. On the other hand, for small black holes, the strings have a dramatic effect, and lead to a second mechanism for small black hole explosion, which is possible even for subcritical 4d potentials. 

The paper is organized as follows. In Section \ref{sec:first-pass} we discuss small black hole explosions in 4d: In section \ref{sec:small-bhs} we introduce small black hole solutions with no 4d potential, in section \ref{sec:perturbing} we derive the equations of motion in the presence of 4d potentials and the critical exponent, and in section \ref{sec:critical-solution1} we construct the critical small black hole solution. In Section \ref{sec:2d-perspective} we approach the systems from the perspective of its truncation to 2d,  performed in section \ref{sec:2d-reduction}. In section \ref{sec:interplay} we describe the interplay of domain walls, black holes and fluxes, and explain the relation between the seemingly different expression for the potentials by using the $\IS^2$ breathing mode. In section \ref{sec:entropy} we adapt the 4d entropy functional formalism to our 2d setup and derive the effective potential for the scalars. In section \ref{sec:effective} we use the effective potential to discuss small black hole explosions and recover the critical exponent. In Section \ref{sec:oned} we discuss running solutions of the 2d theory: In section \ref{sec:running} we use this perspective to recover the equations of motion derived from 4d approaches in the literature. In section \ref{sec:dyn-cob} we describe solutions for small black holes in the subcritical and critical cases. In section \ref{sec:local} we match this picture with the local description of dynamical cobordisms. In Section \ref{sec:obstructions} we describe the consistency conditions requiring certain black holes to emit strings, and their implications for small and large black holes. In section \ref{sec:zk} we discuss dual descriptions of Abelian gaugings and their corresponding symmetries, and the string emission effect. In section \ref{sec:fw} we review this in the illustrative example of the Freed-Witten anomaly. In section \ref{sec:strings-attached} we discuss large black holes in the  4d $\NN=2$ gauged supergravity context, and their emission of strings. In section \ref{sec:banquet} we show that small black holes emitting strings explode, even if the 4d potential is subcritical.
 In Section \ref{sec:cdc} we discuss the ability of small black holes to continue exploring infinite distance limits in moduli space, even in the presence of 4d potentials. After reviewing the situation in exact moduli spaces in section \ref{sec:small-cbc}, we introduce 4d potentials in section \ref{sec:stu} in an illustrative example in the context of the $STU$ model in 4d $\NN=2$ gauged supergravity. This motivates the proposal of a general criterion restricting the properties of general 4d theories in section \ref{sec:constraints}. In Section \ref{sec:conclu} we offer some final thoughts. Appendix \ref{all-h} contains some technical aspects complementing the discussion of Section \ref{sec:first-pass}. Appendix \ref{sec:entropy-4d} provides a derivation of the effective potential of \cite{Bellucci:2008cb} from the entropy function formalism in 4d. In Appendix \ref{again-2d} we provide an alternative rederivation of the equations of motion for running solutions of the 2d theory. Appendix \ref{sec:intro-sugra} collects some basic ingredients of 4d $\NN=2$ (possibly gauged) supergravity.

{\bf Note:} As this work was being completed, we noticed \cite{vandeHeisteeg:2023ubh} which also explores the use of small black holes to constrain the properties of 4d theories near infinite distance limits. It would be interesting to explore the interplay of both approaches to these systems.

\section{Small black hole explosions: the 4d view}
\label{sec:first-pass}

In this section we discuss small black holes and their fate under the introduction of a 4d potential, in a fairly direct 4d description. We follow a formulation similar to \cite{Denef:1998sv,Denef:2000nb} (see also \cite{Hamada:2021yxy} for a recent discussion).

\subsection{Small black holes}
\label{sec:small-bhs}

We start with the structure of small black holes in the absence of 4d potential.
We consider 4d Einstein-Maxwell theory coupled to a real scalar $\phi$:

\begin{equation}
    S= \int d^4x \sqrt{-g} \left[ R- 2 \vert d \phi \vert^2 + \frac{1}{2g^2} \vert F \vert^2 \right]
    \label{action_unper}
\end{equation}

where the gauge coupling is a function $g(\phi)$.

We are looking for spherically symmetric static solution with electric charge $Q$. We take the ansatz
\begin{equation}
    ds^2 = - e^{2U(\tau)} dt^2 + e^{-2 U(\tau)} \left( \frac{d \tau^2}{\tau^4 } + \frac{1}{\tau^2} d \Omega^2_2 \right)
    \label{4dmetric}
\end{equation}

\begin{equation}
    F_2 = 2 \sqrt{2} g^2 Q e^{2U}  d \tau \wedge dt
    \label{ansatz1}
\end{equation}

\begin{equation}
    \phi = \phi (\tau)
    \label{ansatz2}
\end{equation}
Here $d\Omega_2$ is the volume form on a unit $\IS^1$, and $\tau$ runs from the small black hole core (located at $\tau\to -\infty$) to asymptotic Minkowski space (located at $\tau=0$).
In general, one can introduce an additional function $h(\tau)$, but it is possible to set $h=1$, as discussed in Appendix \ref{h-nopot}.

Plugging the ansatz into the action, we get the 1-dimensional action:

\begin{equation}
    S_{1d} = \int d \tau \left\lbrace \dot{U}^2 + \dot{\phi}^2 + 2 g^2 Q^2 e^{2U} \right\rbrace
    \label{1daction}
\end{equation}

This describes the dynamics of one particle in two dimensions under a potential given by (minus) the last term.
The equations of motion read
\beqa
 \ddot{U} = 2 Q^2 e^{2U} g(\phi)^2  \quad,\quad \ddot{\phi}  = Q^2 e^{2U} \left( g(\phi)^2 \right)'       
 \label{eoms}
 \eeqa
where prime denotes derivative with respect to $\phi$.
This must be supplemented with the (`zero energy') constraint
\begin{equation}
     \left[ \frac{1}{2} (\dot{U}^2+ \dot{\phi}^2)\right] -g^2 Q^2 e^{2U} =0
     \label{Hamiltonian_unpert}
\end{equation}
In this work we are going to focus on the particular class of theories with gauge coupling which for $\phi\to\infty$ behave as
\beqa
g=e^{-\alpha \phi}
\label{the-gauge-coupling}
\eeqa
This includes theories arising from 4d $\NN=2$ supergravity. Moreover, the exponential dependence is the behaviour required by the distance conjecture, see \cite{Hamada:2021yxy} for a recent discussion. It would be straightforward to carry out the  analysis of our work for other dependencies, which are left as an exercise for the interested reader.

In this case, the equations of motion read
\beqa
\ddot{\phi}&=& -2\alpha Q^2 e^{2U-2\alpha\phi}\nonumber \\
\ddot{U}&=& 2Q^2 e^{2U-2\alpha\phi}
\eeqa
The combination $\alpha U+\phi$ does not appear in the potential, and corresponds to free motion. Using the initial conditions  $U(\tau=0)=0$ (so that we recover asymptotic flat space) and $\phi(\tau=0)=\phi_0$ (the asymptotic vev at an arbitrary value in the scalar moduli space), we have
\beqa
\alpha U+\phi = \phi_0 + v\tau
\label{free1}
\eeqa
Let us go for the combination that does appear in the potential. For shorthand notation, we denote 
\beqa
\varphi\equiv U-\alpha \phi\quad ,\quad q^2=2(1+\alpha^2) Q^2
\label{def-varphi}
\eeqa
The equation $\ddot{\varphi}= q^2 e^{2\varphi}$ admits the solution
\beqa
\varphi=-\log(-q\tau+c)\, ,
\eeqa
where the initial conditions at $\tau=0$ implies that the constant $c=e^{\alpha\phi_0}$.

The solution is thus
\beqa
U-\alpha\phi & = & -\log(-q\tau+e^{\alpha\phi_0})\nonumber\\
\alpha U+\phi&=& \phi_0
\label{the-sol}
\eeqa
where we have already set $v=0$ in (\ref{free1}), as this turns out to be required by the constraint (\ref{Hamiltonian_unpert}). This gives
\beqa
e^{-2U}&=& e^{-\frac{2\alpha}{1+\alpha^2}\phi_0}\, (-q\tau + e^{\alpha\phi_0})^{\frac 2{1+\alpha^2}} \label{the-u}\nonumber\\
\phi&=&\frac{1}{1+\alpha^2}\phi_0+\frac{\alpha}{1+\alpha^2}\log(-q\tau+e^{\alpha\phi_0})
\label{small-bh-sol}
\eeqa
The region $\tau\to -\infty$ (the small black hole core) is a point at finite spacetime distance, at which the scalar goes off to infinite distance with a logarithmic profile, and the size of the $\IS^2$, given by $e^{-2U}/\tau^2$ in (\ref{4dmetric}), goes to zero.  It is important to notice that this local behaviour is independent of the asymptotic value of $\phi$. As will be clear in later sections, this is the small black hole version of the attractor mechanism.

For completeness we also see that at $\tau\to-\infty$, $g$ tends to zero as
\beqa
g\sim \tau^{-\frac{\alpha^2}{1+\alpha^2}}
\eeqa

\subsection{Perturbing the small black hole with a 4d potential}
\label{sec:perturbing}

We now consider introducing a 4d potential $V(\phi)$ in the theory. The action reads
\begin{equation}
    S= \int d^4x \sqrt{-g} \left[ R- 2 \vert d \phi \vert^2 + \frac{1}{2g^2} \vert F \vert^2 + 2V \right]
    \label{action_pert}
\end{equation}

 The discussion of $h$ now is more involved, but, as explained in Appendix \ref{h-yespot}, one can show that fairly generically $h\simeq 1$ near $\tau\to-\infty$. Since we are interested in the fate of the small black hole core, we simplify the discussion and just proceed with $h=1$. The more careful analysis will be accounted for in the explicit solutions in the next sections.
 
The dynamics is described by the following 1-dimensional action:

\begin{equation}
    S_{1d} = \int d \tau \left\lbrace  \dot{U}^2 + \dot{\phi}^2 + 2 g^2 Q^2 e^{2U}-  e^{-2U} \tau^{-4} V\right\rbrace
 \label{v-tot}
\end{equation}

The 4d scalar potential can modify the solution substantially. For instance, near $\tau=0$, the second term dominates, and the scalar will be fixed at the minimum of the 4d potential (if it exists\footnote{We  assume that such minimum exists, and consider small black holes whose asymptotic vev $\phi_0$ is precisely that minimum. The fate of the small black hole core will nevertheless be independent of this assumption.}). On the other hand, near $\tau\to-\infty$, the black hole potential term would seem to dominate and maintain the basic features of the small black hole: a scalar going off to infinity at the black hole core. This is indeed the case unless the 4d potential grows too fast near $\phi\to\infty$ (i.e. near infinity in field space), as we explore now in more detail.

In order to proceed, we need to parametrize the asymptotic behaviour of the 4d scalar near $\phi\to\infty$. As discussed in \cite{Angius:2022aeq}, a well-motivated proposal is an exponential\footnote{We note that this $\delta$ differs from that in \cite{Angius:2022aeq} by a factor of $\sqrt{2}$, due to different normalization of the scalar kinetic term.}
\beqa
V= V_0 e^{\delta \phi}\quad ,\quad V_0>0
\label{the-v}
\eeqa
We emphasize that we focus on potentials growing at infinity $\phi\to\infty$, namely $V_0>0$ and $\delta>0$, so that they have a chance of competing with the black hole and prevent the scalars from reaching off to infinity.
As we will see, (\ref{the-v}) is also a natural parameterization to make an easy comparison with the black hole pull on scalars, controlled by the exponential gauge coupling (\ref{the-gauge-coupling}).

We can now check when the last term in (\ref{v-tot}) becomes relevant as compared with that  for a small black hole solution, by simply evaluating them on the small black hole solution (\ref{small-bh-sol}). We have that, as $\tau \to-\infty$,
\beqa
&&g^2 Q^2 e^{2U}=Q^2 e^{2(U-\alpha\phi)}\sim 
\tau^{-2}\nonumber\\
&& \tau^{-4}e^{-2U}V\sim 
\tau^{-2+\frac{\alpha(\delta-2\alpha)}{1+\alpha^2}}
\label{the-second}
\eeqa
Hence, for the two terms to be comparable as $\tau \to-\infty$, we get the criticality condition
\beqa
\delta=2\alpha
\eeqa
which remarkably relates the behaviour of the 4d potential and the gauge kinetic functions near infinity in field space.

For $\delta<2\alpha$, the 4d potential becomes subdominant near $\tau\to-\infty$, so the small black hole behaviour prevails, and the small black hole remains small. On the other hand, for $\delta>2\alpha$, the 4d potential is too strong near the putative small black hole core, so that it is energetically too costly to maintain the small black hole behaviour; the small black hole must explode into a finite size solution, either in a runaway towards larger and larger sized, or to a puffed-up finite size black hole whose horizon scalar value is located in a regime where both terms can attain a balance. Which of these two possibilities is realized depends on details beyond the simple exponential approximation to the 4d potential and gauge kinetic function.

We will see examples in later sections.

\subsection{Solution for the critical case}
\label{sec:critical-solution1}

Let us consider the critical case $\delta=2\alpha$. The equations of motion are
 \beqa
&& h_0^2 \ddot{U}=2Q^2 e^{2\varphi}+\frac {V_0 }{\tau^4}e^{-2\varphi}\nonumber \\
&& h_0^2 \ddot{\phi}=-2\alpha Q^2e^{2\varphi}-\alpha\frac{V_0}{\tau^4}e^{-2\varphi}
\label{eoms-crit}
 \eeqa
where $h_0$ is a constant discussed in Appendix  \ref{h-yespot}, and we have defined
  \beqa
  \varphi=U-\alpha\phi
  \eeqa
  There is one freely moving combination
  \beqa
\alpha\ddot{U}+\ddot{\phi}=0\quad \to \quad \alpha U  +\phi =0
  \eeqa
 (with hindsight we have chosen integration constants equal to zero, since $\phi_0$ is irrelevant for $\tau\to -\infty$, and the velocity is eventually set to zero by the hamiltonian constraint).
 
The orthogonal combination $\varphi$ obeys
 \beqa
\ddot{\varphi}=q^2 e^{2\varphi}+v_0\tau^{-4}e^{-2\varphi}
\label{eom-pot}
\eeqa
with
\beqa
  q^2=\frac{2Q^2(1+\alpha^2)}{h_0^2}\quad ,\quad v_0=\frac{V_0(1+\alpha^2)}{h_0^2}
  \label{def-q-v}
  \eeqa 

We now explore the existence of small black hole solutions, by using the following profile for $\varphi$
\beqa
\varphi=-\log(-A\tau)
\label{trial-pot}
\eeqa
Plugging the trial function (\ref{trial-pot}) in the eom (\ref{eom-pot}), we have
\beqa
\frac{1}{\tau^2}=\frac{q^2}{A^2\tau^2}+ \tau^{-4}v_0A^2\tau^2
\eeqa
So we get a solution if
\beqa
v_0 A^4-A^2+q^2=0
\eeqa
This is bi-quadratic. We solve for $A^2$ as
\beqa
A^2=\frac{1- \sqrt{1-4q^2v_0}}{2v_0}
\label{the-a}
\eeqa
where we choose the negative signs because it reproduces the usual small black hole when $v_0$ is very small. We then have
\beqa
U=-\frac{1}{1+\alpha^2}\log(-A\tau)\quad , \quad
\phi=\frac{\alpha}{1+\alpha^2}\log(-A\tau)
\label{sol-crit}
\eeqa
Notice that this is compatible with actually having constant $h$ in (\ref{seico}), 
\beqa
h_0=1+V_0A^2
\eeqa
for a suitable choice of $h_0$ (determined by combining the equation above with (\ref{the-a}) and using the definitions (\ref{def-q-v})). It is also straightforward to compute that the constraint in the last equation in (\ref{constraint-with-h}) is solved.

Since we are interested in $v_0>0$, we get the constraint $q^2v_0<1/4$. This means that even black holes of moderate charge admit no solution, hence cannot remain small, but rather explode. In later sections we will describe tools characterizing the ultimate fate of the former small black holes. In particular, in the critical case it is possible that the black holes are stabilized at a finite horizon puffed up black holes, as we will see in later sections\footnote{In addition to the 4d $\NN=2$ examples discussed in Section \ref{sec:obstructions}, see \cite{Anabalon:2013qua,Anabalon:2013sra} for examples of puffed up black holes in a phenomenological bottom up approach. It is easy to check they correspond to the critical case.}.

\section{Small black hole explosions: The 2d view}
\label{sec:2d-perspective}

In this Section we describe small black holes and their possible explosions from the perspective of the 2d theory resulting after reduction on the angular $\IS^2$. 
Note that, because the compactification scale of the $\IS^2$ will ultimately be comparable with the energy scales of the 2d theory this is intended as a consistent truncation rather than a reduction to a 2d effective field theory.

\subsection{The 2d reduction}
\label{sec:2d-reduction}

In this section we consider the 2d theory resulting from compactification of the 4d theory on $\IS^2$, with gauge fluxes\footnote{We focus on solutions with $\IS^2$ horizons, although in  asymptotically AdS spaces there can be horizons with zero or negative curvature.}.

Consider the following 4d action for a set of complex scalars $z^i$ coupled to a set of Abelian electric and magnetic gauge bosons \footnote{As compared with section \ref{sec:first-pass}, we have changed the sign of the gauge kinetic terms to adapt it to the compactification below, and the sign of the scalar potential to adapt to complex fields and explore asymptotic regions towards $z^i \mapsto -i \infty$.}
\begin{equation}
    S_{4d} = \int d^4x \sqrt{-g_4} \left\lbrace R_4 -2 g_{i \bar{\jmath}} \partial_{\mu} z^i \partial^{\mu} \bar{z}^{\bar{\jmath}} - \frac{1}{2} {\rm Im}\, \mathcal{N}_{\Lambda \Sigma} F^{\Lambda}_{\mu \nu} F^{\Sigma \mu \nu} - \frac{1}{2}{\rm Im}\, \mathcal{N}^{\Lambda \Sigma} G_{\Lambda \mu \nu} G_{\Sigma}^{ \mu \nu} - 2 V \right\rbrace .
    \label{action-again}
\end{equation}

Here $\mathcal{N}_{\Lambda \Sigma}$ and $ \mathcal{N}^{\Lambda \Sigma}$ are the (possibly non-diagonal) electric and magnetic couplings,  $F^\Lambda$ and $G_\Lambda$ are the electric and magnetic field strengths, and $V$ is a general 4d scalar potential.
 
This action is general, but we maintain a notation adapted to future use in 4d $\NN=2$ examples.

\noindent
We perform a compactification on $\IS^2$ with the ansatz:
\beqa
  &&  ds^2_4 =   ds^2_2 + e^{2 \sigma} d \Omega^2_2\nonumber  \\
  & &F^{\Lambda} = \sqrt{2} p^{\Lambda} \sin \theta d \theta \wedge d \varphi\quad ,\quad 
  G_{\Lambda}= \sqrt{2} q_{\Lambda}  \sin \theta d \theta \wedge d \varphi
 \label{ansatz:4d}
\eeqa
 where $p^{\Lambda}$ and $q_{\Lambda}$ are respectively the electric and magnetic charges\footnote{Note that this ansatz is implicitly consistent with electric/magnetic duality, with the proviso that  one introduces additions 2d spacetime filling components of the fields strengths.}
Also, the 2d spacetime dependent warp factor $e^{2 \sigma}$ controls the size of the $\IS^2$. 
The reduced 2-dimensional action is (after removing an overall $4\pi$ factor):

\beqa
    S_{2d} =  \int d^2x \sqrt{-g_2} e^{2\sigma} & \left[ R_2 - 4 \Delta \sigma - 6 \left( \partial \sigma \right)^2 - 2  g_{i \overline{j}} \partial_{\mu} z^i \partial^{\mu} \overline{z}^{\overline{j}} \right. 
    & \left. + 2 e^{-2 \sigma } -  2 e^{-4 \sigma} V_{BH}  - 2 V  \right]\nonumber 
    \eeqa

Where we have introduced the black hole potential 
\begin{equation}
    V_{\rm BH}= - \frac{1}{2} \left( p^{\Lambda} {\rm Im}\, \mathcal{N}_{\Lambda \Sigma} p^{\Sigma} + q_{\Lambda} {\rm Im}\, \mathcal{N}^{\Lambda \Sigma} q_{\Sigma} \right)
    \label{bh-potential}
\end{equation}

Integrating by parts the term $\Delta \sigma (r)$, we get an additional contribution for its kinetic term, giving:
\begin{equation}
    S_{2d} =  \int d^2x \sqrt{-g_2} e^{2 \sigma} \left[ R_2 +2 \left( \partial \sigma \right)^2 - 2 g_{ij} \partial z^i \partial \overline{z}^{\overline{j}} + 2  e^{- 2  \sigma} - 2  e^{-4 \sigma} V_{BH}   - 2 V  \right]
     \label{2d-action}
\end{equation}

\subsection{Interplay of black holes, fluxes and domain walls}
\label{sec:interplay}

Let us make a small aside in this section, which is independent from the general discussion.

A good motivation for the 2d perspective is that it puts both the 4d potential and the black hole potential on a similar footing, namely, they both contribute to the potential of the 2d theory. A particularly nice realization of this is in the context of 4d $\NN=2$ theories, see Appendix \ref{sec:intro-sugra} for notation and definitions, as we discuss next. 

In 4d $\NN=2$ gauged supergravity, the scalar potential can be regarded as arising from (possibly generalized) fluxes in the compactification. These can be regarded as having been produced by domain wall, carrying (3-form) charges $g_\Lambda, g^\Lambda$ whose central charge ${\cal L}$ gives the superpotential (\ref{n2-supo})
\beqa
{\cal L}=e^{{\cal K}/2}\langle {\cal G},v\rangle=e^{{\cal K}/2}(Z^\Lambda g_\Lambda-\partial_\Lambda Fg^\Lambda)
\label{supo-dw}
\eeqa
 leading to a 4d scalar potential (\ref{4d-potential})
\beqa
 V=g^{i{\bar\jmath}} \,{\cal D}_i {\cal L}\,{\bar{\cal D}}_{\bar\jmath}{\bar{\cal L}}-3|{\cal L}|^2
\label{4d-potential-dw}
\eeqa
where we recall that ${\cal D}_if=\partial_if+\frac 12(\partial_i{\cal K}\cdot f)$

On the other hand, black holes with charges $q_\Lambda,q^\Lambda$ arise from 4d particles characterized by a central charge (\ref{n2-central})
\beqa
{\cal Z}=e^{{\cal K}/2}(Z^\Lambda q_\Lambda-\partial_\Lambda F q^\Lambda)
\eeqa
The black hole potential (\ref{bh-potential}) is then given by
\beqa
V_{BH}=g^{i{\bar\jmath}} \,{\cal D}_i {\cal Z}\,{\bar{\cal D}}_{\bar\jmath}{\bar{\cal Z}}+|{\cal Z}|^2
\label{bh-potential-dw}
\eeqa

The identical structure of the central charges ${\cal L}$ and ${\cal Z}$ arises basically from the fact that, in the microscopic description, both the domain walls and the 4d particles are obtained from the same kind of object in what respects to the internal geometry. For instance, in type IIB compactifications, domain walls arise from D5-branes wrapped on 3-cycles of the internal CY, while particles arise from D3-branes wrapped on 3-cycles as well. Hence, the corresponding central charges are controlled by the same geometric quantities. Clearly, we have a similar picture for the (mirror) type IIA compactifications, where domain walls arise from D2-, D4-, D6, D8-branes wrapped on 0-, 2-, 4-, 6-cycles, while 4d particles arise from D0-, D2-, D4-, D6-branes on the same kind of even-dimensional cycles.

On the other hand, there is a seemingly puzzling difference in the expression of the corresponding potentials $V$ and $V_{BH}$. This was recently considered in \cite{Cribiori:2023swd}, but here we propose a more natural explanation. The expression for the potentials differ because they correspond to potentials in theories of different dimensionality. However, in the context of our $\IS^2$ compactification, both 4d domain walls and 4d particles compactify onto 2d particles (or 2d domain walls, since in 2d both kind of objects are the same). Hence in this context, the contribution of both potentials must be on equal footing, as we saw in (\ref{2d-action}), and morally, given by the 2d formula $V=|{\cal D}{\cal Z}|^2+|{\cal Z}|^2$, where here ${\cal Z}$ means the central charge of either kind of object.

The point is that, the different microscopic origin of $V$ and $V_{BH}$ manifests in their different coupling with the new field $\sigma$ arising from the $\IS^2$ compactification. The presence of this extra field in the theory accounts for the factor of $+4|{\cal L}|^2$ necessary to turn (\ref{4d-potential-dw}) into the form (\ref{bh-potential-dw}). The argument is as follows\footnote{ This is analogous to how in 4d $\NN=1$ `no-scale' models,  the scalar potential $|{\tilde  D}W|^2-3|W|^2$ turns into $| DW|^2$ (where the $\tilde D$ in the first expression includes the K\"ahler moduli and in the second does not).}

Consider complexifying the $\IS^2$ breathing modulus (e.g. by including the NSNS 2-form background in type II models) into a new complex modulus $z=b+ie^{-\sigma}$, and let us introduce a K\"ahler potential
\beqa
k(z,{\ov z})=-4\log (-i(z-{\ov z}))
\label{kahlercito}
\eeqa
Note that this reproduces the kinetic term of $\sigma$ in (\ref{2d-action}).

There is thus a total K\"ahler potential given by
\beqa
{\tilde{\cal K}}=k(z,{\ov z})+{\cal K}
\eeqa
where the last piece is the usual 4d K\"ahler potential for vector multiplet moduli. In what follows, we use a tilde to denote quantities defined with this modified K\"ahler potential. For instance, this also modifies the K\"ahler derivatives to
\beqa
{\tilde D}_if=\partial_i f+\frac 12\partial_i{\tilde {\cal K}}\cdot f=\partial_i f+\frac 12\partial_i(k+{\cal K})\cdot f
\eeqa

Let us now consider the modified version of the superpotential (\ref{supo-dw}). We have
\beqa
{\tilde{\cal L}}=e^{{\tilde{\cal K}}/2}\langle {\cal G},v\rangle= e^{k/2}{\cal L}
\label{new-l}
\eeqa
The 2d potential will be given by the tilded version of (\ref{4d-potential-dw})
\beqa
 V=|{\tilde D}{\tilde{\cal L}}|^2-3|{\tilde{\cal L}}|^2
\eeqa
Computing the derivatives
\beqa
{\tilde D}_z{\tilde{\cal L}}=
\frac{4}{(z-{\ov z})^3}{\cal L}\quad ,\quad
{\tilde D}_i{\tilde {\cal L}}
=e^{k/2}D_i{\cal L}
\eeqa
we easily get
\beqa
 V= g^{z{\ov z}}{\tilde D}_z{\tilde{\cal L}} {\ov{\tilde  D}}_{\ov z}{\ov{\tilde{\cal L}}}+ e^{k}g^{i{\ov j}}D_i{\cal L}{\ov D}_{\ov j}{\ov{\cal L}}-3e^k|{\cal L}|^2
= |D{\tilde{\mathcal{L}}}|^2+|{\tilde{\mathcal{L}}}|^2
\eeqa
As promised, we recover an expression identical to the black hole potential (\ref{bh-potential-dw}), by simply replacing the corresponding central charges.

Let us finish by mentioning another argument for the fact that the 4d superpotential must be appropriately dressed with the $\IS^2$ breathing mode to becomes {\em on par} with the 2d black hole central charge. From our above discussion, the natural covariantly holomorphic quantity in the 2d theory is a combination of ${\cal Z}$ and the dressed superpotential (\ref{new-l}), 
\beqa
e^{k/2}{\cal L}=-\frac 1{(z-{\ov z})^2}{\cal L}\sim e^{2\sigma} {\cal L}
\eeqa
This has precisely the required structure of the holomophic superpotential  (\ref{bps-flow}) introduced in \cite{DallAgata:2010ejj}.

\subsection{The entropy function and the effective potential}
\label{sec:entropy}

In this section we use the 2d action (\ref{2d-action}) to obtain the conditions holding in the near horizon limit of a general regular black hole solution. The computation turns out to be a 2d version of the entropy functional formalism computation, see Appendix \ref{sec:entropy-4d} for the 4d computation, following \cite{Sen:2005wa} (see \cite{Sen:2007qy} for a review).

We consider an AdS$_2$ ansatz\footnote{We focus on this, rather than flat of dS$_2$, at least for the time being.} for the metric
\begin{equation}
    ds_2^2 = v_1 \left( -r^2 dt^2 + \frac{dr^2}{r^2} \right)     \label{2d_metric}
\end{equation}
We also take a constant value for the moduli $z$, and denote the size of the $\IS^2$ by $v_2$
\beqa
e^{2 \sigma}=v_2
\eeqa
Evaluating (\ref{2d-action}) on this solution, we get the quantity
\beqa
     \mathcal{E} (v_1,v_2,p,q)=- 4 \pi \sqrt{-g_2} \mathcal{L}_2  
     = 8 \pi v_2 - 8 \pi v_1 + 8 \pi V_{BH} \frac{v_1}{v_2} + 8 \pi V v_1 v_2 
\label{2d-functional}
\eeqa
where the factor of $4\pi$ is just to meet the conventions of the entropy functional formalism. Note that in this formalism, as compared with Appendix \ref{sec:entropy-4d}, it is not necessary to perform the Legendre transform, since the magnetic charges were included in the original action.

The condition to get AdS$_2$ vacua corresponds to the minimization of (\ref{2d-functional}) with respect to the variables $v_1$, $v_2$ and the moduli $z$. We have
\beqa
      & \textit{(i)} \quad \quad  \frac{\partial \mathcal{E}}{\partial v_1} =0 \quad,\quad&  V v_2^2 -v_2 +V_{BH} =0 
       \nonumber \\
       & \textit{(ii)} \quad \quad   \frac{\partial \mathcal{E}}{\partial v_2} =0\quad,\quad& 1 - V_{BH} \frac{v_1}{v_2^2} + V v_1 =0 
       \nonumber\\
       & \textit{(iii)}\quad  \frac{\partial \mathcal{E}}{\partial z^i} \Big\vert_{z^i_H} =0 \quad, \quad & \frac{v_1}{v_2} \frac{\partial V_{BH}}{\partial z^i} \Big\vert_{z^i_H} + v_1 v_2 \frac{\partial V}{\partial z^i} \Big\vert_{z^i_H} =0 
    \label{extremal_cond_par}
\eeqa

The first two give
\beqa
v_2 = \frac{1 - \sqrt{1-4V V_{BH}}}{2 V}\quad ,\quad  v_1 = \frac{v_2}{\sqrt{1-4 V_{BH}  V }}
\eeqa

and replacing into the third, we recover the extremization of the quantity (which in fact corresponds to the $\IS^2$ size, hence the horizon area)
\beqa
V_{eff}=\frac{1-\sqrt{1-4V_{BH}V}}{2V}
\label{eff-pot}
\eeqa
This turns out to be the effective potential derived in \cite{Bellucci:2008cb}, in a different approach related to that in section \ref{sec:running}. This is just a reflection of the equivalence of the entropy functional and attractor mechanism.

\subsection{Small black hole explosions from effective potential}
\label{sec:effective}

It is now a simple exercise to exploit $V_{eff}$ to determine the fate of small black holes upon the introduction of a 4d potential. Consider a small black hole with purely electric charge $Q$. We take the $U(1)$ gauge coupling $g(\phi)=e^{-\alpha\phi}$, as in previous discussions (generalization to other possible functional dependencies is straightforward), hence
\beqa
V_{BH}=2 Q^2 e^{-2\alpha\phi}
\eeqa

Since we are interested in the fate of the core of the small black hole, where $\phi\to \infty$, we parametrize the 4d scalar potential by an exponential as in (\ref{the-v})
\beqa
V=V_0e^{\delta\phi}\quad V_0>0
\label{positive-v0}
\eeqa

The effective potential (\ref{eff-pot}) is
\beqa
V_{eff}=\frac{1-\sqrt{1-8V_0Q^2e^{(\delta-2\alpha)\phi}}}{2V_0e^{\delta\phi}}
\eeqa
We can use it to easily recover the different cases introduced in Section \ref{sec:first-pass}:

$\bullet$ If $\delta<2\alpha$, we may expand the square root and get 
\beqa
V_{eff}\simeq 2 Q^2e^{-2\alpha\phi}=V_{BH}
\eeqa
The 4d potential is too weak at the black hole core and the attractor is unchanged. The scalar is dragged to $\phi\to \infty$, and the small black hole remains small. Note that this does not rule out the possible existence of finite size black holes, although this is model dependent as it involves the behaviour of the gauge coupling and 4d potential at finite values of the moduli; hence, the model independent results on which our analysis focuses is that the small black hole persists.

$\bullet$ The critical case $\delta=2\alpha$ leads to
\beqa
V_{eff}=\frac{1-\sqrt{1-8V_0Q^2}}{2V_0}\, e^{-2\alpha\phi}
\eeqa
For $V_0Q^2\leq1/8$, the effective potential is exponentially damped as in $V_{BH}$, but with different prefactor. The scalar is still dragged to $\phi\to \infty$ by the black hole, which remains small. On the other hand, for $V_0Q^2>1/8$, the effective potential is not well defined, meaning that there is no horizon, neither at finite values of the moduli (and black hole size) nor at infinity.  Of course this conclusion is reliable only as long as the approximation of the gauge coupling and 4d potential as simple exponentials holds. If their behaviour away from infinity in moduli space is different, the small black hole may stabilize into a  puffed-up finite size black hole.\\

$\bullet$ When $\delta>2\alpha$, the exponential in the square root blows up and the effective potential is not well defined. As above, the small black hole explodes in a runaway fashion, at least in the region where the exponential approximations to the gauge coupling and 4d potential are reliable.

\section{Small black hole explosions: The Dynamical Cobordism view}
\label{sec:oned}

As explained in the introduction, the $\IS^2$ reduction in the previous Section is the natural setup in which small black hole solutions can be regarded as dynamical cobordisms of the resulting 2d theory, given by the running of the 4d scalar and the $\IS^2$ size along the radial direction. The small black hole core plays the role of the ETW brane in the sense of \cite{Angius:2022aeq}, namely the cobordism defect that removes the fluxes and allows spacetime to end in the radial direction. From this perspective, small black hole explosions signal a failure of the dynamical cobordism, due to growing corrections as one approaches the putative ETW brane; in case a puffed-up black hole solution exists, the running solution describes the eventual stabilization of the running solution in a (previously hidden) AdS$_2$ vacuum.

\subsection{The 1d running system}
\label{sec:running}

In this section we display black holes as running solutions of the 2d theory (\ref{2d-action}) derived in section \ref{sec:2d-reduction}.
For convenience we perform the following redefinition
\beqa
e^{2\sigma}=r^2e^{2C}
\eeqa
The effective action in terms of $C$ reads
\beqa
   && S_{2d} = \int  d^2x \sqrt{-g_2} r^2 e^{2C} \left\{  R_2 - 2 g_{i \overline{\jmath}} \partial_{\gamma} z^i \partial^{\gamma} \overline{z}^{\overline{\jmath}} + 2 \partial_{\gamma} C \partial^{\gamma} C + \frac{4}{r} C' e^{-2B} + \right. \nonumber \\
    &&\quad\quad\quad\quad\quad\quad\quad\quad\quad \left. + \frac{2}{r^2} (e^{-2B}+ e^{-2C}) - 2\frac{V_{BH}}{r^4} e^{-4C} -2 V_{4d} \right\}
    \label{2d-action-withc}
\eeqa

where the prime denotes the derivative respect to the radial coordinate. 

Now we consider the following ansatz for the 2d metric:
\begin{equation}
    ds^2_2 = -e^{2 A} dt^2 + e^{2B} dr^2
\end{equation}
and we assume that the functions appearing in the metric and the 2d scalars $z^i$ and $C$ just depend by the radial coordinate. We also  restrict to the simplest setup of one single real scalar $\phi$ with $g_{\phi\phi}=1$.

Plugging in the ansatze in the 2d action (\ref{2d-action-withc}), we get the following 1d action
\begin{equation}
\begin{split}
    S = \int  dr  e^{A-B+2C} r^2 &\left\lbrace  C'(C'+2A')+\frac2r\left(C'+A'\right)+\frac{1}{r^2}(1+e^{2(B-C)})-\phi'^2+\right. \\
    & \left. - \frac{e^{2B-4C}}{r^4} V_{BH} -e^{2B} V\right\rbrace
\end{split}
\end{equation}
where we have integrated by parts to remove some total derivatives.

This is an effective action controlling the dynamics of running solutions of the 2d theory. The equations of motion from variations of $A,B,C$ and $\phi$, respectively, are\\
\beqa
&& -e^{-2B}\Big[ 2C''+C^{\prime }(3C^{\prime }-2B^{\prime })+\frac{2}{r}(3C^{\prime
}-B^{\prime })+\frac{1}{r^{2}}\left( 1-e^{2(B-C)}\right) +\phi'^2\Big]
=\frac{e^{-4C}}{r^{4}}V_{BH}+V
\nonumber\\
&& -e^{-2B}\Big[C^{\prime }(C^{\prime }+2A^{\prime })+\frac{2}{r}(C^{\prime
}+A^{\prime })+\frac{1}{r^{2}}\left( 1-e^{2(B-C)}\right) -\phi'^2\Big]=\frac{e^{-4C}}{r^{4}}V_{BH}+V\nonumber \\
 &&-e^{-2B}\Big[A^{\prime \prime }+C^{\prime \prime }+A^{\prime }(A^{\prime
}-B^{\prime })+C^{\prime }(C^{\prime }-B^{\prime }+A^{\prime })+\frac{1}{r}%
(A^{\prime }-B^{\prime }+2C^{\prime }) + \phi'^2\Big]=\nonumber\\
&&\quad\quad\quad\quad\quad=-\frac{
e^{-4C}}{r^{4}}V_{BH}+V\nonumber\\
&&-e^{-2B}\Big[2\phi''+2\phi'\left(A^{\prime }-B^{\prime
}+2C^{\prime }+\frac{2}{r}\right)\Big] =-\frac{e^{-4C}}{r^{4}}\frac{\partial V_{BH}}{\partial \phi}-\frac{\partial V}{\partial \phi}
\label{eoms-bfmy1}
\eeqa
The above ansatz and the resulting equations of motion correspond to those in \cite{Bellucci:2008cb} (for the particular case of a single scalar with canonical kinetic term). We have rederived them as controlling the running solution of the 2d theory, triggered by the presence of the non-trivial potential in (\ref{2d-action}), in the spirit of \cite{Buratti:2021yia,Buratti:2021fiv,Angius:2022aeq}. In Appendix \ref{again-2d} we provide an alternative rederivation of these equations of motion.

It is interesting to mention that, as explained in \cite{Bellucci:2008cb}, it is possible to derive the effective potential (\ref{eff-pot}) from these equations, by plugging in an AdS$_2$ ($\times \IS^2$) ansatz with 
\beqa
A=-B=\log \frac{r}{r_A}\quad ,\quad C=\log\frac{r_H}r
\eeqa
with $r_A$, $r_H$ the horizon value of AdS$_2$ and $\IS^2$ radii, respectively. We refer the reader to  \cite{Bellucci:2008cb} for details.

\subsection{Small black hole solutions Redux}
\label{sec:dyn-cob}

In this language, small black hole solutions correspond to dynamical cobordisms, in which the 1d running solution hits a singularity, at which the scalars (including the $\IS^2$ size) go off to infinite distance in field space, with the small black hole core playing the role of cobordism defect. On the other hand, finite size black holes correspond to setups in which the running halts and the 2d theory reaches an AdS minimum.

In the previous section we derived the equations of motion (\ref{eoms-bfmy1}), which control this running. Using reparametrization of $r$, one can set $A=-B\equiv U$. After some algebra, the equations can be recast as
\begin{equation}
\begin{array}{l}
U'^2+\phi'^2+\frac12\left(\psi''+e^{-2\psi}\right)=\frac{e^{-4C-2U}}{r^{4}}V_{BH}; \\
\\
C''+C'^2+\frac{2}{r}C'+\phi'^2=0; \\
\\
\phi''+2\phi'\psi' =\frac12\frac{e^{-4C-2U}}{r^{4}}\frac{\partial
V_{BH}}{\partial \phi}+\frac12e^{-2U}\frac{\partial V}{\partial \phi}; \\
\\
\psi''+2\psi'^2-e^{-2\psi}=-2e^{-2U}V.
\end{array}
\label{eq3}
\end{equation}
where we have introduced $\psi$, defined as
\beqa
C+U\equiv \psi -\log r
\label{def-psi}
\eeqa

It is worthwhile to point out that the functions $B$ and $\psi$ are related to $U$ and $h$ in Section \ref{sec:first-pass} by
\beqa
B=-U\quad ,\quad h(\tau)=\psi'(r)e^\psi\quad ,\quad \tau=e^{-\psi}
\eeqa
In particular the last equation in (\ref{eq3}) becomes equivalent to the hamiltonian constraint from the vanishing of the integrand in (\ref{almost-constraint}).

We now turn to describing diverse small black hole solutions.

\subsubsection{Small black holes with zero or subcritical 4d potential}

As we anticipated in section \ref{sec:first-pass}, the small black hole solution for subcritical 4d potential is just the small black hole solution in the absence of 4d potential, $V=0$. One can check that this corresponds to $C=-U$, hence, from (\ref{def-psi}), we have $\psi=\log r$. The equations of motion \eqref{eq3} read:

\begin{equation}
\begin{array}{l}
U'^2+\phi'^2=\frac{e^{2U}}{r^{4}}V_{BH}; \\
\\
U''+\frac{2}{r}U'-U'^2-\phi'^2=0; \\
\\
\phi''+\frac2r\phi' =\frac12\frac{e^{2U}}{r^{4}}\frac{\partial
V_{BH}}{\partial \phi},
\end{array}
\label{EoM_smallBH}
\end{equation}
We now focus on the behaviour $V_{BH}=Q^2e^{-2\alpha\phi}$ near $r\to0$, and perform a change of coordinates $r\to-1/\tau$ to get
\begin{equation}
\begin{array}{l}
\dot{U}^2+\dot{\phi}^2=Q^2e^{2(U-\alpha\phi)}; \\
\\
\ddot{U}=Q^2e^{2(U-\alpha\phi)}; \\
\\
\ddot{\phi} =-\alpha Q^2e^{2(U-\alpha\phi)},
\end{array}
\label{EoM_smallBH3}
\end{equation}
These are precisely the constraint (\ref{Hamiltonian_unpert}) and the equations (\ref{eoms}).
The resulting solution is therefore 
\beqa
U=
-\frac{1}{1+\alpha^2}\log (-q\tau) \quad , \quad 
\phi=
\frac{\alpha}{1+\alpha^2}\log (-q\tau),
\label{profile_sBH_sub}
\eeqa
where we have dropped some possible integration constants, irrelevant in the near-core $\tau\to-\infty$ region. These are just the near core behaviour of the solution (\ref{small-bh-sol}).

\subsubsection{Small black holes in the Critical case}

The critical case of the small black hole is given when in the near horizon limit ($r\to0$), the $V$ term is not negligible with respect to $V_{BH}$ but comparable. Restoring the $V$ term, the equations of motion \eqref{EoM_smallBH} are:

\begin{equation}
\begin{array}{l}
U'^2+\phi'^2-\psi'^2+e^{-2\psi}=e^{2U-4\psi}V_{BH}+e^{-2U}V; \\
\\
U''+2U'\psi'=e^{2U-4\psi}V_{BH}-e^{-2U}V; \\
\\
\phi''+2\phi'\psi' =\frac12e^{2U-4\psi}\frac{\partial
V_{BH}}{\partial \phi}+\frac12e^{-2U}\frac{\partial V}{\partial \phi}; \\
\\
\psi''+2\psi'^2-e^{-2\psi}=-2e^{-2U}V.
\end{array}
\label{EoM_smallBH_critical}
\end{equation}
Already here, we can understand the critical case as having $C+U={\rm const}\neq 0$. Using the last equation of \eqref{EoM_smallBH_critical}, this implies that $e^{-2U}V=c_0e^{-2\psi}$, which implies $\psi=\log r+\frac{1}{2}\log(1-2c_0)$, where $c_0$ is a constant that satisfies $c_0<\frac{1}{2}$.

Since in the critical case both potentials must give comparable contributions, we have that $e^{2U-4\psi}V_{BH}\sim e^{-2U}V\propto e^{-2\psi}$, which implies that $V_{BH}\propto e^{-2U+2\psi}$. Hence $V\sim 1/V_{BH}$, in agreement with the analysis in terms of $V_{eff}$ in section \ref{sec:effective}. Using now $V_{BH}=Q^2e^{-2\alpha\phi}$, hence $V=V_0e^{2\alpha\phi}$, and introducing the $r=-(\tau\sqrt{1-2v_0})^{-1}$ change of coordinates, we have

\begin{equation}
\begin{array}{l}
\dot{U}^2+\dot{\phi}^2=\frac{1}{1-2c_0}\left[Q^2e^{2(U-\alpha\phi)}-\frac{V_0}{\tau^4}e^{-2(U-\alpha\phi)}\right]; \\
\\
\ddot{U}=\frac{1}{1-2c_0}\left[Q^2e^{2(U-\alpha\phi)}-\frac{V_0}{\tau^4}e^{-2(U-\alpha\phi)}\right]; \\
\\
\ddot{\phi}=\frac{-1}{1-2c_0}\left[\alpha Q^2e^{2(U-\alpha\phi)}-\alpha \frac{V_0}{\tau^4}e^{-2(U-\alpha\phi)}\right].
\end{array}
\label{EoM_smallBH_critical4}
\end{equation}

These are precisely the equations (\ref{eoms-crit}) and the corresponding constraint c.f. (\ref{constraint-with-h}). 
The resulting solution is  
\beqa
U=-
\frac{1}{1+\alpha^2}\log (-A\tau) \quad , \quad \phi=
\frac{\alpha}{1+\alpha^2}\log (-A\tau),
\label{profile_sBH_crit}
\eeqa
with
\beqa
A^2=\frac{1- \sqrt{1-4q^2v_0}}{2v_0}
\label{the-a-bis}
\eeqa
where we have chosen the negative sign, since it reproduces the usual small black hole solution when $v_0$ is very small. Note that we have dropped some possible integration constants which are irrelevant in the near-core $\tau\to-\infty$ region. These are just (\ref{sol-crit}) with $A$ given by (\ref{the-a}).

\subsection{Small black holes and 2d Dynamical Cobordism}
\label{sec:local}

In this section we would like to make more precise the statement that small black hole solutions correspond to a 2d dynamical cobordism. Following \cite{Buratti:2021fiv, Angius:2022aeq}, dynamical cobordisms are spacetime dependent solutions running along one dimension $y$ which hit a singularity at finite spacetime distance, at which scalars run off to infinite field space distance and at which spacetime ends. In \cite{Angius:2022aeq} it was shown that, near this end of the world brane, dynamical cobordism solutions are described by a simple local model, which implies specific universal scaling relations among the spacetime distance $\Delta$ to the singularity, the field theory distance $D$ traversed, the scalar curvature $R$, and the scalar potential. In particular, in our 2d context
\beqa
D\simeq -\frac{2}{\delta_{2d}}\log \Delta\quad ,\quad R\sim \Delta^{-2}\quad ,\quad
V_{2d}(\phi)\simeq e^{\delta_{2d} D}.
\label{Dynamical_cobordism_def}
\eeqa
where we have added a $2d$ subscript to the potential and critical exponent, to distinguish them from their 4d counterparts.

Let us consider the small black hole in the critical case (the case with subcritical (aka zero) 4d potential can be worked out similarly, as in fact done in \cite{Angius:2022aeq}), namely the solution (\ref{profile_sBH_crit}), and the 2d action (\ref{2d-action-withc}). 
The metric is given by:
\begin{equation}
    ds^2=-e^{2U}dt^2+e^{-2U}dr^2
\end{equation}
Using \eqref{profile_sBH_crit}, we have that $R\propto r^{\frac{-2\alpha^2}{1+\alpha^2}}$. For the spacetime distance we get
\begin{equation}
    \Delta\simeq\int e^{-U}dr\propto r^{\frac{\alpha^2}{1+\alpha^2}}
\end{equation}
Hence we recover the scaling $R\propto \Delta^{-2}$ in (\ref{Dynamical_cobordism_def}) near $r=0$.

Consider now the field distance, from the kinetic terms for the  original scalar field and the $S^2$ radion. Using the profiles for the scalars, we obtain $D(r)=\pm\frac{\sqrt{1-\alpha^2}}{(1+\alpha^2)}\log r$, which in terms of the spacetime distance $\Delta$ becomes\footnote{The result below corrects a typo in eq.(5.40) in \cite{Angius:2022aeq}.}.
\begin{equation}
    D\simeq -\frac{\sqrt{1-\alpha^2}}{\alpha^2}\log \Delta\ , \ \ \ \ \ 
\end{equation}
namely the scaling \eqref{Dynamical_cobordism_def} with the critical exponent 
\beqa
\delta_{2d}=\frac{2\alpha^2}{\sqrt{1-\alpha^2}}\, .
\label{delta-2d}
\eeqa

Finally, consider the 2d potential in (\ref{2d-action-withc}). It is easy to check that, using the solution, all terms in the potential scale as $V_{2d}\sim  e^{2\alpha\phi}$. Hence we have 
\beqa
V_{2d}\sim \exp \Big(\,\frac{2\alpha^2}{\sqrt{1-\alpha^2}}D\,\Big)
\eeqa
Hence reproducing the scaling \eqref{Dynamical_cobordism_def} with the critical exponent (\ref{delta-2d}).

\medskip

In the language of dynamical cobordisms, small black hole explosions can be seen as a failure of the dynamical cobordism, which is obstructed by the existence of a too fastly growing 4d potential at infinity in scalar field space. In cases where the small black holes puffs up into a regular black hole, instead of an ETW brane, the 2d running configuration ends up in an AdS$_2$ vacuum corresponding to the near horizon geometry of the finite size black hole. Hence, the failure of the dynamical cobordism can be blamed on the fact that we do not reach a singularity at a finite distance, but rather is replaced by the infinite AdS$_2$ throat. We will comment more on the impact of such missing dynamical cobordisms in section \ref{sec:cdc}.

It would be interesting to extend these ideas to higher-dimensional theories to build non-trivial AdS vacua, possibly near infinity (hence in weakly coupled regimes) in moduli space.

\section{Explosions from topological obstructions}
\label{sec:obstructions}

In this Section we present a different mechanism which can lead to explosions for small black holes, even for subcritical exponential behaviour of the potential. 
It is based on a simple and familiar mechanism, which has nevertheless been gone unnoticed in the literature on (finite size) black holes in 4d $\NN=2$ gauged supergravity.

\subsection{Topological obstructions from gaugings}
\label{sec:zk}

We explain the mechanism in a simplified, but illustrative enough, setup. It amounts to the statement that in Abelian gaugings, when the $U(1)$ gauge symmetry is broken, its magnetic charges are confined. We follow the discussion in \cite{Banks:2010zn}.

Consider a $U(1)$ gauge theory and an axion scalar, which we denote $\varphi$ (we note that the only relation with $\phi$ is that they often appear as axion-saxion partners in complex scalars, e.g. in supersymmetric theories). We ignore gravity in what follows, since it can be included straightforwardly. Sketchily, we start from the action 
\beqa
    S= \int d^4x (|d\varphi|^2 +  |F |^2 )
\eeqa
The theory has a $U(1)$ gauge symmetry, and a global $U(1)$ shift symmetry for $\varphi$ (i.e. a $(-1)$-form global symmetry). Given that the moduli space has a $U(1)$ isometry, we can now define new theories in which the $U(1)$ gauge transformation acts as a shift along it. The resulting action is
\beqa
\int \, d^4x\,(\, |d\varphi-kA_1|+  |F |^2\,).
\label{lagrangiano-Abeliano}
\eeqa
where $k$ is an integer characterizing the winding number of the gauge $U(1)$ around the  $U(1)$  isometry of the scalar.
The gauge transformation is
\beqa
A_1\to A_1+d\Lambda(x) \quad ,\quad \varphi\to \varphi + k\Lambda(x)
\label{gauging1}
\eeqa
If the axion can be regarded as the phase of a charge $k$ complex field, this just corresponds to a Higgs mechanism, in which the gauge $U(1)$ is broken down to $\IZ_k$.

Upon dualizing $\varphi$ into a 2-form $B_2$, the action shows a St\"uckelberg BF coupling 
\beqa
\int \, d^4x\,(\, |d\varphi|^2+kB_2F_2 |H_3|^2\,)\,
\label{bf}
\eeqa
where $H_3=dB_2$. We can regard this action as the gauging of a global 1-form symmetry of electromagnetism, by  coupling the current $j=F_2$ to the $B_2$ gauge field. This can be made more manifest by dualizing the gauge potential into its magnetic potential $V_1$, the action becomes
\beqa
\int\,d^4x\,(\, |dV_1-kB_2|^2 + |H_3|^2\,)
\eeqa
So the gauge symmetry is
\beqa
B_2\to B_2+d\Lambda_1(x)\quad,\quad V_1\to V_1+k\Lambda_1
\label{gauging2}
\eeqa
The gauging has important implications for the structure of observable operators. In particular, charged operators under the $(-1)$-form symmetry of $\varphi$ (instantons), which have the structure $e^{i\varphi(x)}$ in the absence of gauging, are now no longer gauge-invariant under (\ref{gauging1}). They must be dressed by the emission of $k$ electrically charged particles, i.e. operators charged under the $A_1$ (equivalently the 1-form global symmetry whose current is $*F_2$) along semi-infinite lines $L$ ending at $x$, as
\beqa
\exp \,[\, {i\varphi(x)}+k\int_L A_1\,]\quad ,\quad \partial L=x
\eeqa
Similarly, the gauge transformation (\ref{gauging2}) implies that magnetic monopoles, charged under $V_1$ (equivalently line operators along $L$ charged under the 1-form symmetry whose current is $F_2$), must be dressed by the emission of $k$ strings charged under $B_2$ (equivalently, the 2-form global symmetry whose current is $H_3$), along a surface $\Sigma$ whose boundary is $L$
\beqa
\exp\,[\,{i\int_LV_1}+k\int_\Sigma B_2\,]\quad ,\quad \partial \Sigma=L
\eeqa
Hence, magnetic monopoles come with strings attached. There is a topological obstruction to the existence of free magnetic charges.

This {\em gauging} procedure is standard in supergravity, where it takes ungauged to gauged supergravity theories. In this context, the Abelian gauging parameters ($k$ in the above example) are also known as FI terms, since they are related by supersymmetry to BF couplings. Moreover, in 4d $\NN=2$ theories, such gaugings are related to the apperance of a scalar potential, see Appendix \ref{sec:intro-sugra}. We will see more about it later.

\subsection{A microscopic realization: The Freed-Witten anomaly}
\label{sec:fw}

The realization of the above ideas in the context of flux compactifications was discussed in \cite{BerasaluceGonzalez:2012zn}, where a prominent role is played by Freed-Witten type consistency conditions for fluxes and branes\footnote{Actually \cite{Freed:1999vc} considered the case D-branes in the presence of torsion $H_3$. The physical picture for D-branes and general $H_3$ appeared in \cite{Maldacena:2001xj}, and for D$p$-branes and RR field strength flux $F_p$ appeared in \cite{Witten:1998xy}. Still, we stick to the widely used term FW anomaly / consistency condition.}. In the following we describe one such realization for later use.

Consider type IIB compactified on a CY threefold $\IX_6$ and introduce RR 3-form fluxes on a symplectic basis $\{a_{\Lambda}, b^{\Lambda}
\}$ of 3-cycles
\beqa
\int_{a_{\Lambda}} F_3= g^{\Lambda}\quad ,\quad \int_{b^{\Lambda}}F_3= -g_{\Lambda}\quad,\quad g^{\Lambda},g_{\Lambda}\in \IZ
\eeqa
We ignore other possible fluxes (such as NSNS 3-form fluxes), and focus on these.
Consider the electric and magnetic gauge potentials arising from the RR 4-form
\beqa
A^{\Lambda}=\int_{a_{\Lambda}} C_4 \quad ,\quad A_{\Lambda}=\int_{b^{\Lambda}} C_4
\eeqa
The reduction of the 10d Chern-Simons coupling $B_2F_3F_5$ leads to St\"uckelberg couplings for their field strengths $F^{\Lambda}$, $F_{\Lambda}$
\beqa
\int_{4d}\,\int_{\IX_6}\,B_2F_3F_5;\to\; \int_{4d}\,B_2\,(\,g^{\Lambda} F_{\Lambda}+g_{\Lambda}F^{\Lambda}\,).
\eeqa
This implies the gauging of a particular linear combination of $U(1)$'s. Equivalently, the different (electric and magnetic) $U(1)$'s act on the same scalar $\varphi$, the dual of $B_2$. We can introduce the vector of electric and magnetic gauging
\beqa
{\cal G}=(g^{\Lambda};g_{\Lambda})
\eeqa
Consider now a particle carrying charges $q_{\Lambda}$, $q^{\Lambda}$, under $A^{\Lambda}$, $A_{\Lambda}$, respectively. We introduce the charge vector
\beqa
Q=(q^{\Lambda};q_{\Lambda})
\eeqa
This corresponds to D3-branes wrapped on the 3-cycle
\beqa
\Pi_{\rm D3}=\sum_{\Lambda} (q_{\Lambda} a_{\Lambda} +q^{\Lambda} b^{\Lambda})
\label{D3-cycle}
\eeqa
As is familiar, a D$p$-brane wrapped on a $p$-cycle with $k$ units of RR $F_p$ field strength flux is not consistent by itself, due to a worldvolume tadpole \cite{Witten:1998xy}, and must emit $k$ fundamental strings. In our example
\beqa
\int_{\Pi_{\rm D3}}F_3 = q_{\Lambda} g^{\Lambda} - q^{\Lambda} g_{\Lambda}\,\equiv\,-\langle {\cal G},Q\rangle
\eeqa
where $\langle\cdot,\cdot\rangle$ denotes the symplectic pairing. Hence, when $k\equiv -\langle {\cal G},Q\rangle\neq 0$, there must be $k$ fundamental strings emitted by the D3-brane.

There are other interesting related ways to reach this conclusion, as follows. We can regard the $F_3$ flux as having been created by a domain wall given by a D5-brane on the dual 3-cycle
\beqa
\Pi_{\rm D5}=\sum_i (g^{\Lambda} b^{\Lambda}+g_{\Lambda} a_{\Lambda})\, .
\eeqa
Now, consider the D3-brane wrapped on (\ref{D3-cycle}) on the flux-less side of the domain wall, hence with no strings attached. When the D3-branes is moved across the D5-brane domain wall to the fluxed side, there is a string creation effect \cite{Bachas:1997ui,Danielsson:1997wq,Bergman:1997gf} (dual to the Hanany-Witten effect \cite{Hanany1997}) at each intersection of the corresponding 3-cycles. Hence, the D3-brane ends up with $k$ strings attached, with
\beqa
k=\Pi_{\rm D3}\cdot \Pi_{\rm D5}=q_{\Lambda} g^{\Lambda} - q^{\Lambda} g_{\Lambda} = -\langle {\cal G},Q\rangle
\eeqa
In this picture, the strings attached to the D3-branes can be derived from the FW anomaly due to the $F_3$ flux created by the D5-branes (as discussed above), and the strings attached to the D5-brane arise from the FW anomaly created by the $F_5$ flux created by the D3-brane.

It is straightforward to derive a similar picture for other string setups, for instance type IIA CY3 compactifications with $F_0$, $F_2$, $F_4$ and $F_6$ flux. In fact, we would like to emphasize that the effect is present independently of the particular microscopic realization, since it follows from the structure of topological operators in the presence of gauging c.f. section \ref{sec:zk}.

\subsection{4d ${\cal N}=2$ gauged supergravity Black Holes come with strings attached}
\label{sec:strings-attached}

There is a substantial literature on black holes in 4d $\NN=2$ gauge supergravity, including BPS solutions \cite{Cacciatori:2009iz,DallAgata:2010ejj,Hristov:2010ri}, and non-supersymmetric solutions \cite{Klemm:2012yg,Klemm:2012vm,Gnecchi:2012kb} (see \cite{Astesiano:2021hro} and references therein for a recent discussion)\footnote{As in previous sections, we focus on black holes with spherical horizons. In asymptotic AdS vacua, it is possible to also have black holes with zero or negative curvature. Although we do not discuss them, we note that the negative curvature cases also experience the phenomenon explained in this section.}.

Interestingly, there are many examples of regular black holes based on (sometimes BPS) solutions whose existence depends on having a non-zero value of the symplectic product
\beqa
\langle {\cal G},Q\rangle=-k\in\IZ
\label{condition}
\eeqa
From our above discussion, these black holes solutions are not totally complete, due to a subtle lack of gauge invariance which requires the introduction of $k$ strings sticking out of the black hole (equivalently, filling out the AdS$_2$ 2d geometry in the near horizon region). 

These strings are not included in the discussions in the supergravity literature, and the detailed study of their properties and backreaction is beyond the scope of this work. Let us however point out that their effect is very different (and potentially much milder) from the one in small black holes. Indeed, for regular black holes, the value of $V_{BH}$ at the horizon is a finite quantity, which moreover is controlled by $Q^2$, whereas the number $k$ of attached strings does not scale with $Q$, and can in fact remain small (typical values being of order 1). Hence for large black holes with large charges, the introduction of the strings corresponds to a $1/Q^2$ modification of $V_{BH}$, and hence to subleading corrections to the properties of the black hole.

It is interesting that the topological obstruction explained in previous sections provides a complementary explanation for the quantization condition (\ref{condition}); it simply relates to the integer number of strings emitted by the black hole. Also, in the next section we will argue that the fact that regularity of the black hole solutions is naturally related to a non-zero value of $\langle {\cal G},Q\rangle$, is related to a novel mechanism for small black hole explosions.

In section \ref{sec:constraints} we will see another interesting avatar of the constraint in $\langle{\cal G},Q\rangle$, related to the capability of small black holes to explore all possible infinite distance limits in moduli space in the presence of 4d $\NN=2$ gauging potentials.

\subsection{Explosions via string banquets}
\label{sec:banquet}

As we have discussed in section \ref{sec:interplay}, in certain theories (such as 4d $\NN=2$ supergravity, in previous section) the presence of gauging also implies the appearance of a 4d potential $V$ (\ref{4d-potential-dw}). From the above discussion, it is clear that in the presence of a potential arising from a gauging vector ${\cal G}$, a black hole with charge vector $Q$ will emit a number of strings (magnetically charged under the axion being gauged) given by the symplectic pairing $\langle{\cal G},Q\rangle$. This leads to an interesting mechanism for the explosion of small black holes\footnote{Strictly speaking, such small black holes are not gauge invariant as isolated objects, hence we regard them as defined as endpoint configurations of the corresponding strings. It is in this sense that one may ask whether it is dynamically preferred for these configurations to remain small (and hence, continue exploring infinite distance in field space) or puff up into a finite size configuration (or possibly a runaway version thereof)}.

We can use the effective potential approach to address this question. Indeed, if one regards a possibly puffed up finite size version of the charged black hole, its near horizon limit corresponds to an AdS$_2\times \IS^2$ configuration, with the black hole charges realized as fluxes piercing the $\IS^2$. Due to the latter, the $BF$ couplings (\ref{bf}) implied by the gauging lead to a tadpole for $B_2$ in the 2d non-compact dimensions. Cancellation of this tadpole is achieved by the introduction of strings (charged under $B_2$) spanning the AdS$_2$ factor, and located at a point on $\IS^2$. This is just the near horizon version of the emitted strings mentioned above.

It is easy to derive this tadpole in certain microscopic realizations. For instance, in the type IIB setup in section \ref{sec:fw}, it arises from the reduction of the 10d Chern-Simons term
\beqa
\int_{2d}\,\int_{\IX_6\times \IS^2}\,B_2F_3F_5;\to\; \int_{2d}\,B_2\,(\,g^iq_i-g_iq^i\,)\,=\, \int_{2d}\,B_2 \langle {\cal G},Q\rangle.
\eeqa
Hence we need to add $-\langle {\cal G},Q\rangle=-k$ strings filling AdS$_2$.

The presence of these strings modifies the system. A full discussion of this backreaction is beyond the scope of this work, but we can describe the key feature by modifying the structure of the 2d action, by introducing an extra source of tension, localized on the $\IS^2$ and filling the AdS$_2$, namely.
\beqa
2kT_0\int d^2x \sqrt{-g_2} 
\eeqa
with the tension $T_0>0$ and where the factor of $2$ is for convenience. 
As such, they behave as an extra term $2kT_0e^{-2\sigma}$ inside the bracket of the 2d action (\ref{2d-action}), with $T_0$ taken constant for simplicity\footnote{Let us note that by constant, we mean that it does not depend on vector multiplet moduli. This statement however may be model-dependent and depend on the particular setup or microscopic embedding in string theory.}.
Note that a positive tension $T_0$ contributes with same sign as the black hole or 4d potentials, and opposite to the contribution of the  $\IS^2$ curvature.

Let us now consider the fate of the core of former small black holes in the presence of the gauging potential, and thus, of the extra strings. Consider the simplified case of a single scalar. Following the steps in section \ref{sec:entropy}, the structure of the effective potential (\ref{eff-pot}) is 
\beqa
V_{eff}=\frac{(1-kT_0)-\sqrt{(1-kT_0)^2-4V_{BH}V}}{2V}
\eeqa
Note that the string contributes as the tadpole generated by the $\IS^2$ compactification, and they contribute with opposite sign. Moreover, in microscopic realizations, the strings are locally BPS objects (such as the fundamental strings in Freed-Witten anoamlies), and by charge quantization we expect that $T_0=1$. 
Hence, in the presence of such strings, $V_{eff}$ goes negative and according to the solutions in section \ref{sec:dyn-cob}, there is no small hole solution. Remarkably, this conclusion is independent of the value of $\delta$ and how it compares with $\alpha$. Namely, in the presence of topological obstructions which force the introduction of strings sticking out of the small black hole, the small black holes explodes even for subcritical values of the exponent in the 4d potential.

An intuitive interpretation is that small black holes with strings attached tend to feast on them and grow. As usual, our exponential approximations near infinity in field space do not suffice to assess or not the existence of an endpoint for this process, a topic to which we turn in the next section.

Incidentally, we note that certain regular black holes, such us the example in the STU model in \cite{DallAgata:2010ejj}, do not correspond to small black holes even in the absence of 4d potential. In this sense, they cannot be regarded as a puffed up version of small black holes. 

\section{Interplay with the Cobordism Distance Conjecture}
\label{sec:cdc}

\subsection{Small black holes and the Cobordism Distance Conjecture}
\label{sec:small-cbc}

The Cobordism Distance Conjecture proposed in \cite{Buratti:2021fiv} states that in a consistent theory of quantum gravity, every infinite field distance limit can be realized as a dynamical cobordism running into an end of the world brane (possibly in a suitable compactification of the theory). In fact, for 4d theories and upon compactification on $\IS^1$ this encompasses the Distant Axionic String Conjecture in \cite{Lanza:2021qsu}. Similarly, this suggests that small black holes, which are dynamical cobordisms of the 4d theory compactified on $\IS^2$, are a powerful tool to explore infinite distance limits of the theory. In this section we explore to what extent the set of small black holes is rich enough to actually explore all those possible limits, and to what extent the obstructions due to small black hole explosions imply some limitations in this exploration.

A clear restriction in this regard is that small black holes allow to explore infinite distance limits for scalars controlling gauge couplings of the theory. This is implicit in the discussion below. In particular, we will frame the discussion in the context of 4d $\NN=2$ supergravity, which provides a rich arena to carry out the discussion in explicit examples.

We start with the theory in the absence of 4d potentials (see \cite{Gendler:2020dfp} for a related discussion). In this context, the structure of infinite distance limits in the vector multiplet moduli space has been studied in \cite{Grimm:2018ohb,Grimm:2018cpv,Corvilain:2018lgw}. In particular, every infinite distance limit corresponds to a regime in which some gauge coupling goes to zero, hence allowing the construction of small black hole solution exploring that infinite distance limit \cite{Hamada:2021yxy}. Note that in general an infinite distance limit can involve combinations of different moduli to go off to infinity (as classified by the corresponding growth sector), hence the corresponding small black hole must carry the appropriate combination of charges to couple to the gauge coupling becoming small. Hence, this implicitly profits from the Completeness Conjecture, which ensures that for any possible charge vector one may need, there are appropriate charged states in the theory (and which for large charges constitute a small black hole, rather than a fundamental particle).

\subsection{Adding the potential: the $STU$ model}
\label{sec:stu}

\subsubsection{Generalities}

Let us now consider the introduction of a 4d potential. In order to be explicit, we focus on those arising from Abelian gaugings of 4d $\NN=2$ theories, as in the previous section, see Appendix \ref{sec:intro-sugra} for background information. In particular, we can focus on the illustrative example of the $STU$ model. We will subsequently study the extension to more general 4d $\NN=2$ models, in particular covering the moduli space of any CY$_3$ compactification.

We follow closely the conventions in \cite{DallAgata:2010ejj}, and refer the reader to Appendix \ref{sec:intro-sugra} for details.

The model (in a specific symplectic frame) is defined by the prepotential
\begin{equation}
	F = \frac{X^1 X^2 X^3}{X^0}\,.
\end{equation}
We introduce the coordinates $S = X^1/X^0$, $T = X^2/X^0$ and $U = X^3/X^0$, in terms of which the K\"ahler potential is
\begin{equation}
	K = - \log [-i (S - \bar S)(T - \bar T)(U - \bar U)],
\end{equation}
and ${\cal V}$ is given by
\begin{equation}
	{\cal V} = e^{K/2}\,(1, S, T, U, -STU, TU, SU, ST)^T.
\end{equation}
We consider a general set of charges/gaugings, given by
\begin{equation}
    Q =  \left(  p^0 , p^1 , p^2 , p^3 , q_0 , q_1 , q_2 , q_3  \right)^T \quad \quad \quad     \mathcal{G} = \left(  g^0 , g^1 , g^2 , g^3 , g_0 , g_1 , g_2 , g_3  \right)^T
\end{equation}
A rich class of models was introduced in \cite{DallAgata:2010ejj} by taking gauging charges of the form ${\cal G} = (0,g^1,g^2,g^3,g_0,0,0,0)^T$ and black hole charges $Q = (p^0,0,0,0,0,q_1,q_2,q_3)$. For the following discussion, we prefer to allow for general choices (in principle constrained only by the Freed-Witten like consistency condition discussed in the previous section). 

The black hole central charge (\ref{n2-central}) and 4d superpotential (\ref{n2-supo}) are thus 
\beqa
&& {\cal Z} = e^{K/2} (p^0 STU-p^1 TU-p^2 SU-p^3 ST+q_0+q_1 S+q_2 T+q_3 U)\cr
 &&{\cal L} =  e^{K/2} (g_0+g^0 STU-g^1 TU+g_1 S-g^2 S U+g_2 T-g^3 ST+g_3 U).
\eeqa
Since we are interested in exploring infinite distance limits in moduli space, the asymptotic behaviours are controlled by the imaginary parts of the moduli, hence we replace $S \mapsto i s$, $T \mapsto i t$ and $U \mapsto i u$, with $s,t,u<0$.

From the expressions for the black hole potential (\ref{bh-potential-dw}), namely (\ref{bh-potential}), and the 4d scalar potentials (\ref{4d-potential-dw}), namely (\ref{4d-potential}), we have
\beqa
    &&V_{BH}= \frac{ q_0^2+ q_1^2 s^2+q_2^2t^2+q_3^2u^2+(p^1)^2 t^2u^2 +(p^2)^2 s^2 u^2 + (p^3)^2 s^2t^2+(p^0)^2 s^2t^2 u^2}{-2stu}\nonumber\\
 &&   V=\frac{ 
 -g_0 (g^1 t u+g^2 s u+g^3 s t)-g_1g_2st-g_2g_3 tu-g_1g_3su}{-stu}+\nonumber\\
 && \quad\quad\quad\quad
 +\frac{ (g^0g_1 -g^2g^3)s^2tu+(g^0g_2-g^1g^3) st^2u+(g^0g_3-g^1g^2) stu^2}{-stu}
 \label{general-pots}
\eeqa
It is now a simple exercise to pick sets of gauging/charges which illustrate different behaviours for small black holes. We refrain from the discussion of random examples, but rather focus on a more fundamental question in the next section.

\subsubsection{Small black holes: exploration or explosion}
\label{explorers}

 As explained, a natural question to what extent small black hole explosions may hamper their ability to explore infinite distance limits in a given theory.
 In this section, we carry out this discussion in the $STU$ model, we may consider the simplified setup in which all moduli are identified i.e. scale to the limit in the same way. More general situations will be discussed later on.

For a single modulus case $s=t=u\equiv z$, we have
\beqa
    &&V_{BH}= \frac{ q_0^2+ (q_1^2 +q_2^2+q_3^2)z^2+\,[\,(p^1)^2 +(p^2)^2 + (p^3)^2 \,]\,z^4+(p^0)^2 z^6}{-2z^3}\\
     &&   V=\frac{ 
 (-g_0 (g^1 +g^2 +g^3) -g_1g_2-g_2g_3 -g_1g_3)z^2}{-z^3}+\nonumber\\
 && \quad\quad\quad\quad
 +\frac{ (g^0g_1+g^0g_2+g^0g_3 -g^1g^2-g^2g^3-g^1g^3)z^4}{-z^3}
 \label{pot-onemod}
\eeqa

As explained in section \ref{sec:small-cbc}, we are interested in testing possible limitations in the ability of small black holes to explore infinite distance limits, due to small black hole explosions.

Hence, we focus on 4d potentials with the highest possible value of $\delta$ in the corresponding limit\footnote{Note that we are also not insisting on the 4d theory to admit an asymptotic AdS$_4$ vacuum, since the discussion of the small black hole cores does not require information about the asymptotics, which may well correspond to a 4d running solution.}. Considering the limit of large moduli\footnote{We Note that there is a similar limit, corresponding to $z\to 0$, which can be analyzed similarly, with identical results up to reshuffling of terms. This is just a reflection of the duality of the $STU$ model.}, the most dangerous term is the second line in (\ref{pot-onemod}), which scales as $V\sim z$. There are basically two possibilities (or simple combinations thereof):

$\bullet$ We may choose $g^0\neq0$, and get $V\sim z$ by choosing some of the $g_i\neq 0$. In order to explore this infinite distance limit, one may be tempted to just take a black hole with $q_0\neq 0$ as only charge, so that $V_{BH}\sim z^{-3}$, for which the 4d potential is subcritical. However, this is not possible, since it would lead to Freed-Witten-like anomalies. However, it is possible to choose black holes with some $q_0=0$, $q_i\neq 0$ for $i=1,2,3$, which avoid the Freed-Witten-like anomalies, and lead to $V_{BH}\sim z^{-1}$, for which the 4d potential is critical. 

$\bullet$ We may get $V \sim z$ by keeping $g^0=0$ and turning on two different $g^i$'s, for instance $g^1$ and $g^2$. In such case, we can take a black hole with $q_1, q_2\neq 0$, with lead to $V_{BH}\sim z^{-1}$, hence a critical case, and which easily avoid the Freed-Witten-like anomalies by demanding that  $g^1 q_1 + g^2 q_2=0$ (e.g. $q_1=g^2 n$ , $q_2= -g^1 n$, with arbitrary $n\in \IZ$).

We have thus checked that, even for the most dangerous 4d potential, there are specific small black holes for which the 4d potential is critical; hence they in principle have a chance to remain small and be able to explore this infinite distance limit. We thus conclude that, also in the presence of 4d potentials, the Completeness Conjecture is enough to guarantee that, for any gauging potential and any infinite distance limit, there is always some small black hole in the theory which remains small and still allows to explore the infinite distance limit. 

We finish by noting that we have not discussed other possible sources of 4d potentials, which in principle could lead to a modification of the above discussion. On the other hand, turning things around, the above discussion motivates a  proposal restricting the form of the possible 4d potentials, as we discuss in the next section.

\subsection{Constraints on potentials from small black holes explosions}
\label{sec:constraints}

In this section we propose the use of our ideas above to constrain the possible 4d scalar potentials, from their properties regarding small black holes and their explosions. We show that this can lead to powerful restrictions, not only on 4d potentials but (in theories with extended supersymmetry) also on the geometry of the moduli space near infinity.

\subsubsection{The Small Black Hole Criterion}

In the example of the $STU$ model in the previous section, we showed the interesting fact that the available 4d potentials from gaugings are such that there still remain enough small black holes (i.e. those possibly not exploding under the effect of the potential) to still allow the exploration of all infinite distance limits for the scalars.

It is natural to propose promoting this idea to a general principle, as follows:\\

{\bf Small Black Hole Criterion:} {\em In a consistent theory of quantum gravity with scalar-dependent Abelian gauge couplings, the allowed 4d potentials are constrained by the requirement that they allow for the existence of a set of small black holes whose scalars are still able to explore any infinite distance limit of the theory (i.e. remain subcritical or critical).}\\

In a sense, this idea is not a new conjecture, but rather a mere reformulation of the Cobordism Distance Conjecture in \cite{Buratti:2021fiv} for the 2d theories resulting from the 4d theories upon $\IS^2$ compactification. On the other hand, its explicit formulation facilitates its use to put constraints on the 4d theories, as we discuss in the next section.

\subsubsection{Recovering the CY vector moduli space asymptotics}

In this section we put the above proposal to work and show that it leads to remarkably strong results. In particular, we will argue it implies that, for any 4d $\NN=2$ theory, the K\"ahler potential for the vector multiplets satisfies, in the large moduli limit, the equality
\beqa
g^{i{\bar {\jmath}}}\partial_i {\cal K}\partial_{\bar{\jmath}} {\cal K}=3
\label{cy-magic}
\eeqa
This is precisely the asymptotic identity obeyed by vector moduli in a CY threefold compactification of type II string theory (and via duality, of K3 compactifications of heterotic, or other dual realizations of 4d $\NN=2$). We are however about to derive it from a purely 4d bottom-up perspective.

The argument is as follows. Consider a general 4d $\NN=2$ with prepotential $F(X)$. Using homogeneity of the prepotential, the covariantly holomorphic vector ${\cal V}$ always contains an entry equal to $e^{{\cal K}/2}$, its symplectic dual contains and entry $e^{{\cal K}/2}F(z)$, where $z$ are the moduli (i.e. affine coordinates). Let us focus on the limit of all moduli becoming large, in which this entry dominates the dynamics. For instance, from (\ref{kpot}) the K\"ahler potential scales as
\beqa
{\cal K}\sim -\log[i(F-{\bar F})]
\eeqa
Consider now a possible gauging of the theory, analogous to $g^0$ in the $STU$ model in section \ref{sec:stu}, namely
\beqa
{\cal L}\sim g^0 F(z) +\ldots
\eeqa
The corresponding scalar potential (\ref{4d-potential-dw}), namely (\ref{4d-potential}), is
\beqa
V=\frac{1}{i(F-{\bar F})}\,\big[\, (\,g^{i{\bar{\jmath}}}\partial_i {\cal K}\partial_{\bar{\jmath}} {\cal K}-3\,)\, (g^0)^2\, F{\bar F}
\big]
\eeqa
which scales as $V\sim F$ in the large moduli limit unless (\ref{cy-magic}). We now show that this would be in contradiction with our Small Black Hole Criterion, hence (\ref{cy-magic}) must be satisfied.

Consider the above potential and demand that there must exist some black holes which is critical with respect to that term in the potential, namely $V_{BH}\sim F^{-1}$. Clearly, this must correspond to a black hole whose only charge, $q_0$,  is associated to the entry $e^{{\cal K}/2}$ in ${\cal V}$, namely the symplectic dual to $g^0$. This would superficially allow to satisfy the Small Black Hole Criterion. However, we should realize that precisely this required charge vector suffers the problem pointed out in Section \ref{sec:obstructions}, namely
\beqa
\langle {\cal G},Q\rangle=g^0q_0\neq 0
\eeqa
Namely, it is not an allowed state in the theory, as it is forbidden by gauge invariance. Alternatively, even if the black hole is made consistent by allowing it to emit the required number of strings, it does not remain small, but rather explodes by the string banquet mechanism in section \ref{sec:banquet}, and does not explore the infinite distance limit. Hence, as announced, the Small Black Hole Criterion implies the property (\ref{cy-magic}) in the infinite moduli limit of the $\NN=2$ vector moduli space of any consistent theory. 

The above argument is a general description of how the $(g^0)^2$ term does not arise in the potential (\ref{general-pots}) of the $STU$ model in section \ref{sec:stu}. In fact this model is one of the simplest toy models to characterize the large moduli behaviour of any CY compactification 
\beqa
F=-D_{IJK} X^IX^JX^k/X^0
\eeqa
with $6D_{IJK}$ are the familiar (integer) triple intersection numbers of CY compactifications.

It is an amusing fact that we recover the very familiar property (\ref{cy-magic}) of CY$_3$ compactifications in a 4d theory which has a priori no information about CY's or compactifications. In this sense, the result resonates with the reconstruction of compactification geometries from cobordism considerations in \cite{Hamada:2021bbz}.

\section{Conclusions}
\label{sec:conclu}

In this work we have explored the ability of 4d small black holes to explore infinite distance limits in scalar field space in the presence of scalar potentials growing near infinity. We have derived the critical relation distinguishing when the small black hole remains small from when it explodes by puffing up into a regular black holes or following a runaway behaviour. Although the discussion is general we have studied the particular case of 4d $\NN=2$ gauged supergravity, which allows a rich arena for this exploration. In fact, in that context we have uncovered a constraint, previously unnoticed in the literature on regular black holes in this setup, which requires the black holes to emit strings. 

We have used these findings to motivate a general constraint on scalar potentials in 4d theories, based on the principle that the set of small black holes should still be rich enough to allow for exploration of all possible infinite distance limits. We have shown that this leads to non-trivial constraints, and that in 4d $\NN=2$ theories it allows to reproduce some of the properties of vector moduli spaces of CY$_3$ compactifications.

Our discussion has emphasized the use of a 2d description, after truncation on $\IS^2$. This allowed us to discuss 4d potentials and black hole potentials on an equal footing, and, in the 4d $\NN=2$ context, to satisfactorily explain the difference between their expressions in terms of the underlying covariantly holomorphic quantities. More importantly, the 2d perspective permits the translation of the result to the language of dynamical cobordisms, where the small black hole core corresponds to the end of the world brane, and small black hole explosions signal the failure of completing a cobordism ending spacetime. In this picture, puffed up black holes correspond to running solutions of the 2d theory which end up relaxing onto a newly appeared AdS$_2$ vacua.

There are many possible generalizations of our work and other related ideas. For instance:

$\bullet$ We expect that further results about the structure of infinite distance regimes in moduli space can be derived for further exploration of large and small black holes, in particular considerations about their entropy and its microscopic explanation, possibly along the lines of \cite{vandeHeisteeg:2023ubh}.

$\bullet$ Generalization to black strings and other extended objects, in general dimensions, and their possible explosions upon the introduction of scalar potentials. A particular instance of this is the fate of the 4d EFT strings in \cite{Lanza:2020qmt,Lanza:2021qsu,Marchesano:2022avb} in the presence of potentials growing fast at the string core.

$\bullet$ The failure of 2d running solutions to complete dynamical cobordisms and rather end up relaxing onto AdS$_2$ vacua is a particularly rich instance of a phenomenon which also exists in higher dimensions \cite{Angius:2022aeq}. It would be interesting to explore its possible use as a technique to build or detect AdS vacua hidden near infinity in moduli space.

$\bullet$ In the supercritical case, the small black hole explosion is expected to lead to time dependent runaway configurations. The construction of such full solution is presumably difficult, due to non-trivial dependence in two coordinates. However, its asymptotic behaviour at late times/distances may admit a simpler description, depending only on a single light-cone coordinate, in analogy with expanding domain walls in e.g. \cite{Hellerman:2004qa,Hellerman:2006nx, Hellerman:2006hf,Hellerman:2006ff,Hellerman:2007fc,Angius:2022mgh}.

$\bullet$ The Freed-Witten-like constraint in 4d $\NN=2$ gauged supergravity models forces the introduction of strings, which in the near horizon regime of regular black holes are spacetime filling in the 2d AdS$_2$. In cases where this extra ingredient breaks supersymmetry i.e. is an anti-string (see \cite{Sugimoto:1999tx,Antoniadis:1999xk,Aldazabal:1999jr}, also \cite{Angelantonj:1999ms,Angelantonj:2000xf,Rabadan:2000ma} for similar phenomena in higher dimensions), this is reminiscent of the antibrane uplift in \cite{Kachru:2003aw}. It would be interesting to explore actual quantitative connections between the two phenomena.

We hope to come back to these and other interesting topics in future work.

\section*{Acknowledgments}

We are pleased to thank Gianguido Dall'Agata, I\~naki Garcia-Etxebarria, Luis Ib\'a\~nez, Fernando Marchesano, Miguel Montero, Tom\'as Ort\'{\i}n, Irene Valenzuela for useful discussions, and Jos\'e Calder\'on-Infante and Matilda Delgado for collaboration on related topics. This work is supported through the grants CEX2020-001007-S and PID2021-123017NB-I00, funded by MCIN/AEI/10.13039/501100011033 and by ERDF A way of making Europe. The work by R.A. is supported by the grant BESST-VACUA of CSIC. The work by J. H. is supported by the FPU grant FPU20/01495 from the Spanish Ministry of Education.

\appendix

\section{All about $h$}
\label{all-h}

\subsection{No 4d potential}
\label{h-nopot}

The most general ansatz for an spherically symmetric static solution with electric charge $Q$ in the theory described by (\ref{action_unper}) is actually
\begin{equation}
    ds^2 = - e^{2U(\tau)} dt^2 + e^{-2 U(\tau)} \left( \frac{d \tau^2}{\tau^4 h(\tau)^2} + \frac{1}{\tau^2} d \Omega^2_2 \right)\quad ,\quad F_2= 2 \sqrt{2} Q g^2 e^{2U} d \tau \wedge dt
    \label{4d-ansatz-h}
\end{equation}

Plugging the ansatz in the 4d action (\ref{action_unper}) leads to the 1d action
\begin{equation}
    S_{1d} = \int d \tau \left\lbrace \frac{h}{2} \left( \dot{U}^2 + \dot{\phi}^2 \right) + \frac{1}{h} g^2 Q^2 e^{2U} - \frac{1}{\tau^2} (\frac{1}{h} -h + 2 \tau \dot{h})\right\rbrace
    \label{1daction}
\end{equation}
The equations of motion read
\beqa
    && \frac{d}{d \tau} \left( h \dot{U} \right) = \frac{2}{h} Q^2 e^{2U} g(\phi)^2   \label{h-dep_EoM1} \\
    && \frac{d}{d \tau} \left( h \dot{\phi} \right) = \frac{1}{h} Q^2 e^{2U} \left( g(\phi)^2 \right)'   \label{h-dep_EoM2}\\
 &&  \left[ \frac{1}{2} (\dot{U}^2+ \dot{\phi}^2) - \frac{1}{\tau^2}\right] h^2 - g^2 Q^2 e^{2U} + \frac{1}{\tau^2} =0\,.   \label{h-dep_EoMh}
\eeqa
As mentioned in \cite{Denef:1998sv}, the action (\ref{1daction}) implies that $h$ is not an actual dynamical variable, but rather imposes a constraint, associated to reparametrizations of $\tau$. To make this explicit, consider the variation of the action under changes of $U$, $\phi$ of the form
\beqa
 U(\tau + \delta \tau) = U(\tau) + \dot{U}(\tau) \delta \tau \quad, \quad
    \phi (\tau + \delta \tau) =  \phi (\tau) + \dot{\phi}(\tau) \delta \tau\,.
    \label{variations}
\eeqa
Using (\ref{h-dep_EoM1}), (\ref{h-dep_EoM2}) we get:
\begin{equation}
    \delta S_{1d} = \int d \tau \left\lbrace \frac{d}{d \tau} \left( h \dot{U}^2 \delta \tau \right) + \frac{d}{d \tau} \left( h \dot{\phi} \delta \tau \right) + \frac{2}{\tau^3}  \left( \frac{1}{h} - h \right) \delta \tau \right\rbrace,
\end{equation}
namely, up to total derivatives, demanding $\delta S_{1d}=0$ requires
\beqa
    \int d \tau \frac{2}{\tau^3} \left( \frac{1}{h} - h \right) \delta \tau = const
\eeqa
which can be solved by setting $h=1$. Then (\ref{h-dep_EoM1}), (\ref{h-dep_EoM2}) correspond to the equations of motion (\ref{eoms}), and  (\ref{h-dep_EoMh}) leads to the constraint (\ref{Hamiltonian_unpert}).

\subsection{Including 4d potential}
\label{h-yespot}

Let us now introduce a 4d potential term in the action c.f. (\ref{action_pert})
\begin{equation}
    S=\int d^4x\sqrt{-g}\left[R-2(d\phi)^2 + \frac{1}{2g^2}F^2+ 2V\right].
\end{equation}
Using the ansatz (\ref{4d-ansatz-h}), we get the 1d action
\beqa
    S_{1d} = \int d \tau \left\lbrace \frac{h}{2} \left( \dot{U}^2 + \dot{\phi}^2 \right) + \frac{1}{h} Q^2 g (\phi)^2 e^{2U} - \frac{1}{h} \frac{V (\phi)}{2 \tau^4} e^{-2U} - \frac{1}{2 \tau^2} \left( \frac{1}{h}- h + 2 \tau \dot{h} \right) \right\rbrace. \nonumber
    \label{1daction_pert}
    \eeqa
The equations of motion are
\beqa
            & &\frac{d}{d \tau} \left( h \dot{U} \right) = \frac{2}{h} Q^2 g(\phi)^2 e^{2U} + \frac{1}{h} \frac{V (\phi)}{ \tau^4} e^{-2U}\nonumber \\
            && \frac{d}{d \tau} \left( h \dot{\phi} \right) = \frac{1}{h} Q^2 \left( g(\phi)^2 \right)' e^{2U} - \frac{1}{h} \frac{V' (\phi)}{2 \tau^4} e^{-2U}, \nonumber\\
           && \left[ \frac{1}{2} \left( \dot{U}^2 + \dot{\phi}^2 \right) - \frac{1}{2\tau^2} \right] h^2 - g(\phi)^2 Q^2 e^{2U} + \frac{V}{2 \tau^4} e^{-2U} + \frac{1}{2\tau^2} =0
        \label{constraint-with-h}
 \eeqa
The change of the action under variations (\ref{variations}) requires
    \begin{equation}
      \delta S_{1d}=  \int d \tau \frac{1}{\tau^3} \left( \frac{1}{h} -h+2 \tau \dot{h} + \frac{1}{h} \frac{ 2V (\phi)}{ \tau^2} e^{-2U} \right) \delta \tau =0 .
      \label{almost-constraint}
    \end{equation}
which can be solve by choosing
    \begin{equation}
     \dot{h} \simeq 0 \quad \quad \quad h^2 = 1 + \frac{ 2 V (\phi)}{ \tau^2} e^{-2U}
    \label{seico}
    \end{equation}
Although this is more involved, it simplifies in particular cases. For instance, if $V e^{-2U}$ grows slower that $\tau^2$ we recover $h^2=1$ near $\tau\to -\infty$, just like in the case with no 4d potential; this corresponds to the subcritical case in the main text. On the other hand, the critical case solution in section \ref{sec:critical-solution1} corresponds to a constant $h\equiv h_0\neq 1$. 
        
\section{The 4d entropy functional computation of $V_{eff}$}
\label{sec:entropy-4d}

For completeness we include a quick derivation of $V_{eff}$ from the entropy functional from the perspective of the 4d solutions. The 2d computation of $V_{eff}$ in section \ref{sec:entropy} is a version of this.

We start with the following 4d action\footnote{This action is equivalent to the 4d action (\ref{action-again}) used at the beginning of section \ref{sec:2d-reduction}, where we made explicit the democratic treatment of the electric and magnetic gauge fields. Here it is the Legendre transformation (\ref{entropy_funct_0}) with the condition (\ref{e_conj}) that will restore the not-manifest electric/magnetic duality.  }
\begin{equation}
    S_{4d} = \int d^4x \sqrt{-g_4} \left\lbrace  R_4 - 2 g_{i \overline{j}} \partial_{\mu} z^i \partial^{\mu} \overline{z}^{\overline{j}} +  Im \mathcal{N}_{\Lambda \Sigma} F^{\Lambda}_{\mu \nu} F^{\Sigma \mu \nu} - 2 V \right\rbrace  
    \label{4d_action}
\end{equation}

Recall that, although it has the structure of the bosonic sector of $\mathcal{N}=2$ (possibly gauged) supergravity action, but is intended as a general theory.

\noindent
We are interested in the near horizon limit of extremal black hole solutions with spherical horizon, which corresponds to AdS$_2\times\IS^2$ with a 2-form field strength background, of the form:
\begin{equation}
    ds_4^2 = v_1 \left( -r^2 dt^2 + \frac{dr^2}{r^2} \right) + v_2 \left( d \theta^2 + \sin^2 \theta d \varphi^2 \right)
    \label{4d_metric}
\end{equation}

\begin{equation}
    F^{\Lambda}_{rt}=\frac{e^{\Lambda}}{8 \pi} \quad \quad \quad F^{\Lambda}_{\theta \phi}= p^{\Lambda} \sin \theta
    \label{F2_form}
\end{equation}
where $v_1, v_2, e$ and $p$ are constants. In addition we require that the scalars be regular when we approach the horizon: $\lim_{r \mapsto r_H} z^i = z^i_H$. \\
Following \cite{Sen:2005wa} (see \cite{Sen:2007qy} for a review), we introduce the \textit{entropy function}, defined by
\begin{equation}
    \mathcal{E} (v_1, v_2, z^i, p, q) = 2 \pi \left( q_{\Lambda} e^{\Lambda} - f(v_1, v_2, z^i, p,e) \right)
    \label{entropy_funct_0}
\end{equation}
where $q$ is the conjugate quantity to $e$:
\begin{equation}
    \frac{\partial \mathcal{E}}{\partial e} =0
    \label{e_conj}
\end{equation}
and  $f$ is the action evaluation on the near horizon ansatz solution
\begin{equation}
    f(v_1, v_2, z^i,  p, e) = \int d \theta d \varphi \sqrt{-g_4} \mathcal{L}_4.
    \label{f_def}
\end{equation}
The parameters for the actual solution, including the attractor values for the scalars, are obtained by extremizing the entropy function
\beqa
        \textit{(i)} \quad \frac{\partial \mathcal{E}}{\partial v_1} =0 \quad ,\quad
         \textit{(ii)} \quad \frac{\partial \mathcal{E}}{\partial v_2} =0 \quad ,\quad
         \textit{(iii)} \quad \frac{\partial \mathcal{E}}{\partial z^i} \Big\vert_{z^i_H} =0 
    \label{E_extremal_cond}
\eeqa
Moreover, the value of $\mathcal{E}$ at this extremum gives the black hole entropy. 

Using the ansatzs \eqref{4d_metric} and \eqref{F2_form} 
the entropy functional \eqref{entropy_funct_0} reads:
\begin{equation}
\begin{split}
    & \mathcal{E} (v_1, v_2, z^i_H, p,q) = 2 \pi \left\lbrace q_{\Lambda}e^{\Lambda} + 8 \pi v_2 - 8 \pi v_1 + \right. \\
    & \left. - 4 \pi Im \mathcal{N}_{\Lambda \Sigma} p^{\Lambda} p^{\Sigma} \frac{v_1}{v_2} + \frac{1}{16 \pi} Im \mathcal{N}_{\Lambda \Sigma} e^{\Lambda} e^{\Sigma} \frac{v_2}{v_1} + 8 \pi V v_1 v_2 \right\rbrace \\
    \end{split}
    \label{E_funct1}
\end{equation}
Here the conjugate variable $q$ is defined in terms of the other parameters by the \eqref{e_conj}:
\begin{equation}
    e^{\Lambda}= - 8 \pi \frac{v_1}{v_2}  Im \mathcal{N}^{\Lambda \Sigma} q_{\Sigma} .
\end{equation}
where $Im \mathcal{N}^{\Lambda \Sigma}$ denotes the inverse matrix of $Im \mathcal{N}_{\Lambda \Sigma}$. \\
Using this last equation, \eqref{E_funct1} becomes:
\beqa
  \mathcal{E} = 2 \pi \left\lbrace 8 \pi v_2 - 8 \pi v_1 - 4 \pi Im \mathcal{N}_{\Lambda \Sigma} p^{\Lambda} p^{\Sigma} \frac{v_1}{v_2} - 4 \pi Im \mathcal{N}^{\Lambda \Sigma} q_{\Lambda} q_{\Sigma} \frac{v_1}{v_2} + 8 \pi V v_1 v_2 \right\rbrace
    \label{E_funct2}
\eeqa
Note that the electric and magnetic charge contributions couple to $v_1$ and $v_2$ in the same way. This motivates introducing the black hole potential:
\begin{equation}
    V_{BH} = - \frac{1}{2} \left( p^{\Lambda} Im \mathcal{N}_{\Lambda \Sigma} p^{\Sigma} +   q_{\Lambda} Im \mathcal{N}^{\Lambda \Sigma} q_{\Sigma} \right)
\end{equation}
Now \eqref{E_funct2} takes the form:
\begin{equation}
    \mathcal{E} (v_1, v_2, p,q) = 2 \pi \left[ 8 \pi v_2 - 8 \pi v_1 + 8 \pi V_{BH} \frac{v_1}{v_2} + 8 \pi V v_1 v_2 \right]
    \label{entropy_funct}
\end{equation}
The extremization conditions \eqref{E_extremal_cond} give:
\begin{equation}
    \begin{split}
       & \textit{(i)} \quad  V v_2^2 -v_2 +V_{BH} =0 \quad \quad \quad \longmapsto \quad \quad \quad v_2 = \frac{1 \pm \sqrt{1-4V V_{BH}}}{2 V} \\
       & \textit{(ii)} \quad 1 - V_{BH} \frac{v_1}{v_2^2} + V v_1 =0 \quad \quad \longmapsto \quad \quad \quad v_1 = \frac{v_2}{\sqrt{1-4 V_{BH}  V }} \\
       & \textit{(iii)} \quad \frac{v_1}{v_2} \frac{\partial V_{BH}}{\partial z^i} \Big\vert_{z^i_H} + v_1 v_2 \frac{\partial V}{\partial z^i} \Big\vert_{z^i_H} =0 \\
    \end{split}
    \label{extremal_cond_par}
\end{equation}

These conditions exactly reproduce those in \cite{Bellucci:2008cb}, derived from the attractor mechanism. \\
Replacing the values of $v_1$ and $v_2$, one can define the effective potential 
\beqa
V_{eff}=\frac{1-\sqrt{1-4V_{BH}V}}{2V}
\eeqa
introduced in (\ref{eff-pot}) in the main text.

\section{Alternative derivation of the 1d equations of motion}
\label{again-2d}

In this Appendix we provide an alternative derivation of the equations of motion in section \ref{sec:running}.

We slightly generalize the ansatz (\ref{ansatz:4d}) for the reduction of the 4d metric to 2d:
\begin{equation}
    ds^2_4 = e^{\alpha \sigma} ds^2_2 + e^{\beta \sigma} d \Omega^2_2
\end{equation}
The resulting 2d action, after integrating by parts some terms, is:
\begin{equation}
\begin{split}
    S_{2d} =  \int d^2x \sqrt{-g_2} e^{\beta \sigma (r)} & \left\lbrace R_2 - \beta \left( \alpha + \frac{\beta}{2} \right) \left( \partial \sigma \right)^2 - 2  g_{i \overline{j}} \partial_{\mu} z^i \partial^{\mu} \overline{z}^{\overline{j}} \right. \\
    & \left. + 2 e^{(\alpha-\beta) \sigma (r)} - 2  V_{BH} e^{( \alpha -2 \beta) \sigma (r)}  - 2 V_{4d} e^{\alpha \sigma (r)} \right\rbrace \\
    \end{split}
    \label{2d_comp_action}
\end{equation}
One can then obtain the equation of motion for this action upon variations of the 2d metric, $\alpha$, $\beta$, $\sigma$ and the moduli $z^i$.

One can then use the following 2d ansatz
\begin{equation}
    ds^2_2 = -dt^2 +e^{2 b(r)} dr^2.
\end{equation}
The relation with the quantities $A,B,C$ in the main text is
\begin{equation}
    e^{\alpha \sigma (r)} = e^{2 A(r)} \quad \quad \quad e^{\alpha \sigma (r) +2 b(r)}= e^{2 B(r)} \quad \quad \quad e^{\beta \sigma (r)} = r^2 e^{2 C(r)}.
    \label{ident_ansats}
\end{equation}
Upon setting, without loss of generality, $\beta =2$, we recover the equation of motion (\ref{eoms-bfmy1}).

\section{4d $\NN=2$ supergravity }
\label{sec:intro-sugra}

A general setup in which 4d potentials are naturally included is gauged 4d $\NN=2$ supergravity. Moreover this provides a template for flux compactifications. Here we review its basic ingredients, see \cite{Lauria:2020rhc} for a thorough discussion. 

\subsection{Ungauged 4d $\NN=2$ supergravity}
\label{sec:ungauged}

We start with the ungauged 4d $\NN=2$ coupled to $n_V$ abelian vector multiplets, and ignore any hypermultiplets in the discussion. There are $n_V$ complex moduli $z^i$ labelled by $i=1,\ldots, n_V$. Including the graviphoton, there are $n_V+1$ gauge bosons, labelled by $\Lambda = 0,\ldots,n_V$. The structure of the bosonic lagrangian is
\beqa
{\mathscr L} = R - 2g_{i\bar\jmath}\, \partial_\mu z^i \partial^\mu \bar z^{\bar \jmath} + \hbox{Im}{\cal N}_ {\Lambda\Sigma}\, F^\Lambda_{\mu \nu}\, F^{\Sigma\,\mu\nu} +\hbox{Re}{\cal N}_{\Lambda\Sigma}\, F^{\Lambda}_{\mu\nu}\,\frac{\epsilon^{\mu\nu\rho \sigma}}{2\sqrt{-g}} F^\Sigma_{\rho \sigma} 
\eeqa
where the different quantities are defined using special geometry. The scalars parametrize a special K\"ahler moduli space, i.e. is the base of a symplectic bundle, with covariantly holomorphic sections
\beqa
{\mathcal V}=\begin{pmatrix} X^{\Lambda}\\ F_{\Lambda} \end{pmatrix}\quad ,\quad {\cal D}_{\bar \imath}{\cal V}=\partial_{\bar \imath}{\cal V}+\frac 12 \partial_{\bar \imath}{\cal{K}}\,{\cal V}=0,
\eeqa
(where ${\cal K}$ is the K\"ahler potential) and obeying the symplectic constraint
\beqa
i\langle {\cal V},{\bar {\cal V}}\rangle=F_{\Lambda}{\bar X}^{\Lambda}-X^{\Lambda}{\bar F_{\Lambda}}=1
\label{constraint-curlyv}
\eeqa
It is useful to write
\beqa
{\cal V}=e^{{\cal K}(z,{\bar z})/2} v
\eeqa
The holomorphic symplectic vector
\beqa
v(z)=\begin{pmatrix} Z^{\Lambda}\\ \partial_{\Lambda} F \end{pmatrix}\quad ,\quad \partial_{\bar i}v=0
\eeqa
is defined using the prepotential\footnote{Although there are 4d $\NN=2$ theories with no prepotential, we focus on the usual case in which it exists.} $F(X)$, a holomorphic function of degree 2. Then, (\ref{constraint-curlyv}) becomes
\beqa
e^{-{\cal K}}=i\langle v,{\bar v}\rangle
\label{kpot}
\eeqa
The matrix ${\cal N}_{\Lambda \Sigma}$ determining the coupling between the scalars  and the vectors is defined by the relations
\beqa
F_{\Lambda}={\cal N}_{\Lambda \Sigma} X^{\Sigma}\quad ,\quad {\cal D}_{\bar \imath}{\bar F}_{\Lambda}={\cal N}_{\Lambda \Sigma}{\cal D}_{\bar \imath} {\bar X}^{\Sigma}
\eeqa

For completeness, we present the microscopic realization of this in type II compactification on a Calabi-Yau threefold $\IX_6$. Start with type IIB, where the vector moduli space corresponds to the complex structure moduli. Introduce a symplectic basis of 3-cycles $A_{\Lambda}$, $B^{\Sigma}$ (namely we have intersection numbers $A_{\Lambda} \cdot A_{\Sigma} =B^{\Lambda}\cdot B^{\Sigma}=0, A_{\Lambda}\cdot B^{\Sigma}=\delta_{\Lambda}^{\Sigma}$). The holomorphic vector $v$ is given by 
\beqa
Z^{\Lambda}=\int_{A_{\Lambda}}\Omega\quad,\quad \partial_{\Lambda} F=\int_{B^{\Lambda}}\Omega
\label{periods}
\eeqa
and 
\beqa
K=-\log\big(\,i\int_{\IX_6}\Omega\wedge {\bar \Omega}\big)
\eeqa

The electric and magnetic gauge potentials arise from the redution of the 10d RR 4-form
\beqa
A_1^{\Lambda}=\int_{A_{\Lambda}} C_4\quad ,\quad A_{1\,\Lambda}=\int_{B^{\Lambda}} C_4
\label{iib-gauge-fields}
\eeqa

For type IIA one has a similar story with even-dimensional cycles replacing the 3-cycles and $e^J$ playing the role of $\Omega$.

\subsection{Gauged 4d $\NN=2$ supergravity}
\label{sec:gauged}

In this section we review some useful formulas for Abelian gaugings in supergravity. Gaugings are defined by introducing a vector of parameters (FI terms)
\beqa
{\cal G}=\begin{pmatrix} g^{\Lambda} \cr g_{\Lambda} \end{pmatrix},
\eeqa
corresponding to electric or magnetic gaugings, respectively. The physical interpretation of the gauging is reviewed in section \ref{sec:zk} in the main text.

The gauging introduces a 4d superpotential
\beqa
{\cal L}=\langle {\cal G},{\cal V} \rangle =e^{{\cal K}/2}(Z^{\Lambda}g_{\Lambda}-\partial_{\Lambda} Fg^{\Lambda})
\label{n2-supo}
\eeqa
This leads to a 4d scalar potential\footnote{Notice that this reproduces the usual formula of 4d $\NN=1$ with the $e^{\cal K}$ prefactor arising from squaring the $e^{{\cal K}/2}$ in the defining of ${\cal V}$, manifest in the last expression in (\ref{n2-supo}).}
\beqa
V=g^{i{\bar\jmath}} \,{\cal D}_i {\cal L}\,{\bar{\cal D}}_{\bar\jmath}{\bar{\cal L}}-3|{\cal L}|^2
\label{4d-potential}
\eeqa

The intuition is very simple. The gaugings correspond to the introduction of fluxes in the compactification (including field strength fluxes, but also geometric or generalized fluxes). This can be described in terms of a domain wall connecting the vacuum of the ungauged theory (i.e. no fluxes) and the vacuum of the theory with gauging ${\cal G}$. The domain wall is a codimension 1 object in 4d, and its internal structure is such that its tension is given by (\ref{n2-supo}). This is the standard argument for the superpotential from $p$-form field strength fluxes in M-theory or type II compactifications \cite{Gukov:1999ya}, which for completeness we repeat for type IIB 3-form fluxes \cite{Taylor:1999ii}. 

Consider type IIB on a CY3 $\IX_6$ as described around (\ref{periods}) and introduce RR 3-form fluxes
\beqa
\int_{A_{\Lambda}}F_3=g^{\Lambda}\quad ,\quad \int_{B^{\Lambda}}F_3=g_{\Lambda}
\label{iib-flux-quanta}
\eeqa
For clarity here and in what follows we ignore constant factors. 

Denoting the basis of 3-forms $\alpha_{\Lambda}$, $\beta^{\Sigma}$ Poincar\'e dual  to $A_{\Lambda}$, $B^{\Sigma}$, the flux quanta (\ref{iib-flux-quanta}) give the 3-form flux cohomology class
\beqa
[F_3]= g^{\Lambda}\alpha_{\Lambda}+g_{\Lambda}\beta^{\Lambda}
\eeqa
This flux can be regarded as being created by a domain wall, given by a D5 wrapped on the 3-cycle $\Pi_{\rm D5}$ whose Poincar\'e dual is
\beqa
\delta(\Pi_{\rm D5})=F_3= g^{\Lambda} \alpha_{\Lambda} +g_{\Lambda} \beta^{\Lambda}
\eeqa
The tension of this domain wall can be obtained as
\beqa
e^{\cal K}\int_{\Pi_{\rm D5}}\Omega=e^{\cal K}\int_{\IX_6} \Omega\wedge \delta(\Pi_{\rm D5}) =e^{\cal K}(Z^{\Lambda} g_{\Lambda}-\partial_{\Lambda}F g^{\Lambda})=e^{{\cal K}/2}\langle {\cal G},v \rangle={\cal L}
\eeqa

\subsection{Black hole central charge}

In 4d $\NN=2$ theories there are BPS particle states, which can correspond to charged black holes. We introduce a vector of gauge charges
\beqa
Q=\begin{pmatrix} p^{\Lambda} \cr q_{\Lambda} \end{pmatrix},
\eeqa
They have an associated central charge
\beqa
{\cal Z}=\langle Q,{\cal V}\rangle=e^{{\cal K}/2}(Z^{\Lambda} q_{\Lambda} -\partial_{\Lambda} Fp^{\Lambda})
\label{n2-central}
\eeqa
The black hole potential which controls the radial flow of scalars in the attractor mechanism is given by
\beqa
V_{BH}=g^{i{\bar\jmath}} \,{\cal D}_i {\cal Z}\,{\bar{\cal D}}_{\bar\jmath}{\bar{\cal Z}}+|{\cal Z}|^2
\label{bh-potential}
\eeqa

The intuition is very simple. For instance in the type IIB setup, BPS particles arise from D3-branes wrapped on a (special lagrangian) 3-cycle $\Pi_{\rm D3}$ whose Poincar\'e dual is
\beqa
\delta(\Pi_{\rm D3})=F_3= p^{\Lambda} \alpha_{\Lambda} +q_{\Lambda} \beta^{\Lambda}
\eeqa
in the symplectic basis introduced in the previous section. The BPS mass can be obtained as
\beqa
e^{\cal K}\int_{\Pi_{\rm D3}}\Omega=e^{\cal K}\int_{\IX_6} \Omega\wedge \delta(\Pi_{\rm D3}) =e^{\cal K}(Z^{\Lambda} q_{\Lambda} -\partial_{\Lambda}F p^{\Lambda})=e^{{\cal K}/2}\langle Q,v \rangle={\cal Z}
\eeqa

In the presence of a gauging introducing a potential, the complete attractor flow is determined by a combination of ${\cal Z}$ and ${\cal L}$. In particular, in \cite{DallAgata:2010ejj} it was shown that for BPS solutions the attractor flow is controlled by the superpotential
\begin{equation}
 {\cal W}=e^U|{\cal Z}+ie^{2\sigma}{\cal L}|   
 \label{bps-flow}
\end{equation}
where $U$ and $\sigma$ are functions of the radial coordinate controlling the non-compact 2d geometry and the $\IS^2$ size, respectively

\bibliographystyle{JHEP}
\bibliography{mybib}

\end{document}